\documentclass[%
reprint,
superscriptaddress,
 amsmath,amssymb,
 aps,
]{revtex4-1}
\usepackage[pdftex]{graphicx}
\usepackage{dcolumn}
\usepackage{bm}
\usepackage[colorlinks=true,linkcolor=blue]{hyperref}
\usepackage[utf8]{inputenc}
\usepackage[english]{babel}
\usepackage{xcolor}

\addto\captionsenglish{}

\newcommand{\un}[1]{\,\mathrm{#1}}

\begin{document}

\preprint{APS/123-QED}

\title{Jet effusion from a metal droplet irradiated by a polarized ultrashort laser pulse}

\author{S.~Yu.~Grigoryev}
\email{grigorev@phystech.edu}
\affiliation{Joint Institute for High Temperatures, RAS, 13/2 Izhorskaya st., 125412 Moscow, Russia}
\affiliation{Dukhov Research Institute of Automatics, 22 Sushchevskaya st., 127030 Moscow, Russia}

\author{B.~V.~Lakatosh}
\affiliation{Joint Institute for High Temperatures, RAS, 13/2 Izhorskaya st., 125412 Moscow, Russia}

\author{P.~M.~Solyankin}
\affiliation{Institute on Laser and Information Technologies, Branch of the Federal Scientific Research Center “Crystallography and Photonics” of RAS, Shatura, 140700, Russia}

\author{M.~S.~Krivokorytov}
\affiliation{Institute for Spectroscopy, RAS, 5 Fizicheskaya Street, 142190, Troitsk, Moscow, Russia}

\author{V.~V.~Zhakhovsky}
\email{basi1z@ya.ru}
\affiliation{Joint Institute for High Temperatures, RAS, 13/2 Izhorskaya st., 125412 Moscow, Russia}
\affiliation{Dukhov Research Institute of Automatics, 22 Sushchevskaya st., 127030 Moscow, Russia}

\author{S.~A.~Dyachkov}
\affiliation{Joint Institute for High Temperatures, RAS, 13/2 Izhorskaya st., 125412 Moscow, Russia}
\affiliation{Dukhov Research Institute of Automatics, 22 Sushchevskaya st., 127030 Moscow, Russia}

\author{C.-D. Ohl}
\affiliation{Institute for Physics, Otto von Guerricke University, Universitätsplatz 2, 39106 Magdeburg, DE-Germany}

\author{A.~P.~Shkurinov}
\affiliation{Faculty of Physics and International Laser Center, Lomonosov Moscow State University, 119991 Moscow, Russia}

\author{V.~V.~Medvedev}
\email{medvedev@phystech.edu}
\affiliation{Institute for Spectroscopy, RAS, 5 Fizicheskaya Street, 142190, Troitsk, Moscow, Russia}

\date{\today}

\begin{abstract}
Fragmentation of liquid metal droplets irradiated by linearly and circularly polarized femtosecond laser pulses is observed in our experiment. The obtained shadowgraph snapshots demonstrate that a circularly polarized pulse may produce several randomly-oriented jets effused from the expanding droplets, while a linearly polarized laser pulse generates strictly the cruciform jets. The latter orientation is tied with polarization plane, rotation of which causes rotation of the cruciform jets by the same angle. To shed light on the experimental data we performed molecular dynamics simulation of droplet expansion induced by angle-dependent heating.  

Our simulation shows that the jet directions are determined by an oriented angle-dependent energy distribution within a frontal hemisphere layer of droplet after absorption of linearly polarized light. As a result, the produced flow velocity field guided by surface tension forms two high-speed opposite jets oriented across the electric field vector as in our experiment. 

A shock-wave pulse generated in the frontal layer has angle-dependent amplitude inherited from the oriented energy deposition. The release part of shock pulse produces a cavitation zone nearby the droplet center, and thus an expanding spherical shell is formed from the droplet. The flow velocities within a rearside hemisphere of the shell, produced after reflection of the shock wave from the rear side of droplet, generate two low-speed opposite jets oriented along the electric field vector. Thus we found that the cruciform jets are originated independently from the frontal and rear sides of droplet, and a pair of frontal jets is faster than a pair of rearside jets.

\end{abstract}

%\keywords{molecular dynamics, extreme ultraviolet lithography, shock wave, fragmentation}

\maketitle

\section*{Introduction}
\label{sec:intro}

To control the spatial and temporal characteristics of an impact on materials being processed using ultra-short laser pulses with sub-picosecond duration a detailed understanding of laser-induced high-strain-rate phenomena in targets is required. Ultrafast energy deposition leads to the generation of shock waves in materials, and in spite of their essential role in the subsequent dynamics, the shock-induced fragmentation mechanism of droplet targets is still poorly understood. Development of the technique, which can precisely guide the fragmentation, is crucial for processing the liquid droplets.

Focusing an intense short laser pulse onto a liquid target can lead to the formation of liquid jets. Such jets appear as a direct response to the pressure build-up and shock wave generation induced by heating in a layer of material where the absorbed laser energy is distributed. Adjusting the laser pulse parameters one can obtain the jet shape which is useful for particular applications. Such jets are mainly used in medicine~\cite{hirano2001enhancement, nakagawa2002holmium, ohki2004experimental} and printing (depositing) materials on different surfaces~\cite{duocastella2008jet, unger2011time, brown2011time, brasz2015tilting}.

The hydrodynamic instabilities may also be induced by a laser pulse irradiation~\cite{klein2020drop, thoroddsen2009spray, li2019cavitation, zeng2018jetting, avila2016fragmentation}. Such phenomena are easier to observe when laser pulse is focused onto the transparent droplet target. For example, a nanosecond Nd:YAG laser pulse was focused into the center of a levitating water droplet in the study~\cite{avila2016fragmentation}. Absorption of a laser pulse in the focal spot leads to rapid heating and formation of a vapor bubble inside a droplet. Explosive growth of the bubble results in jetting from the droplet surface~\cite{plesset1954stability, prosperetti1977viscous}.

Hydrodynamic instabilities and jet effusion from liquid droplets are the key processes involved in generation of laser-induced plasma sources for Extreme Ultra-Violet Lithography (EUVL), see details in the recent review~\cite{versolato2022microdroplet}.
Liquid jets formed during the laser-induced fragmentation of a liquid metal droplet are discussed in~\cite{krivokorytov:2018,grigoryev2018expansion}. There the impulsive energy deposition that drives the explosive fluid mechanic is rather different. As it was shown in~\cite{grigoryev2018expansion}, a femtosecond laser pulse irradiation of spherical liquid metal droplet launches a convergent shock wave propagating inwards the droplet. The outgoing shock front reflects as a rarefaction wave that is geometrically focused and for sufficient amplitude ruptures the liquid metal droplet. As a result, a rapidly expanding cavitation bubble is nucleated within the droplet. The reflection of the shock waves and the explosive growth of the cavitation bubble result in a destabilization of the droplet surface into liquid metal jets, which eventually fragment into smaller and fast secondary droplets due to the Plateau-Rayleigh instability.

Previously we studied the importance of the laser energy on details of the jetting ~\cite{krivokorytov:2018}. The important finding was, that with higher deposited energy, higher order surface modes of the droplet are excited. These surface modes determine the number and direction of the liquid jets. In the present article the study is extended to take into account the polarization of laser pulse. It is expected that the polarization affects the spatial distribution of the reflected light and therefore the pressure distribution within shock wave propagating in the droplet~\cite{de2021cylindrically}. Here the linear and circular polarized light are studied experimentally and in atomistic simulation.

The used molecular dynamics (MD) method has the ability to naturally include cavitation through well chosen inter-atomic potentials in contrast to a continuum fluid approach. This benefit is paid off with the computing demand that requires smaller droplets in the MD simulations as compared to the experiments. Through a suitable scaling, see Ref. ~\cite{grigoryev2018expansion}, the experimental and numerical results are compared.  We demonstrate that the experimental results agree well with simulated evolution of irradiated tin droplet.

\section*{Experimental setup}

\begin{figure}[t]
	\flushleft{
		\includegraphics[width=1.0\linewidth]{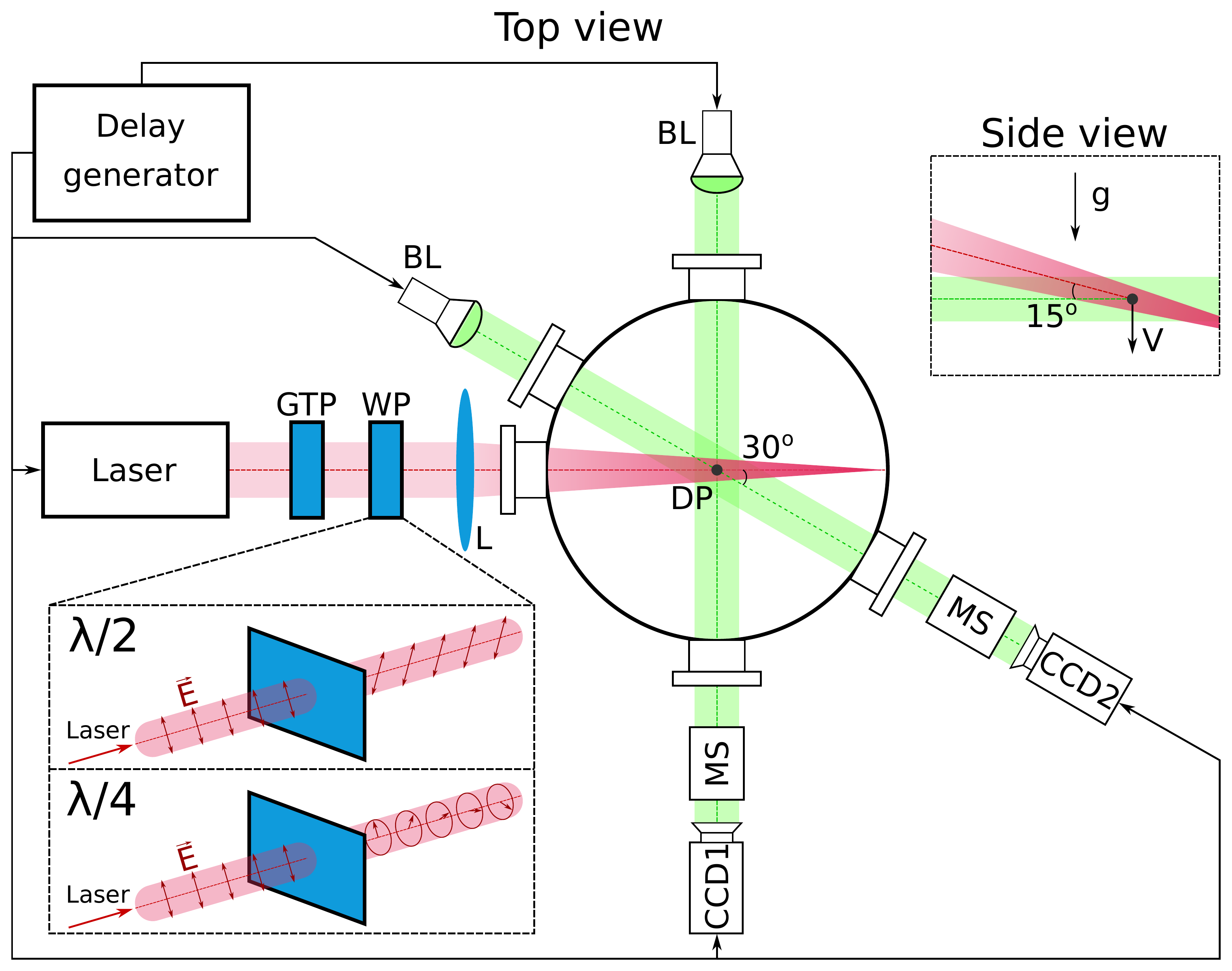} 
		\caption { The scheme of the experimental setup. The setup includes the lens (L); the liquid metal droplet (DP); the back-light laser (BL); the microscope (MS); the cameras CCD1 and CCD2; the Glan--Taylor prism (GTP); the waveplates, which were situated at positions (WP).} \label{fig:exp_setup}
	}
\end{figure}

The scheme of the experimental setup used in our experiments is shown in Fig.~\ref{fig:exp_setup}. Liquid metal droplets with diameter $30 \pm 2\,\mu$m are produced by the previously designed droplet generator~\cite{vinokhodov_dropgen}. We use Sn-In liquid alloy (in mass proportion $48\%-52\%$) for droplets generation. The alloy is heated to $140 ^{\circ}C$ during experiments, well above the melting temperature of the alloy ($119^{\circ}C$), which is almost two times lower than that of pure tin and indium. A relatively low operating temperature simplifies the experiment, keeping all the other physical properties (density, viscosity, and surface tension) of the alloy similar to those of pure tin. During irradiation, the droplet is in free fall ($V \sim 10\,$m/s directed straight down, see Fig.~\ref{fig:exp_setup}) in a vacuum chamber with a residual gas pressure less than $10^{-4}\,$mbar. 

In the experiment a Ti:Sapphire laser (Spectra Physics Spitfire Ti:Sa regenerative amplifier system), operating at $797\,$nm central wavelength, is used to irradiate liquid metal droplets by a focused femtosecond pulse. The used laser pulse energy of $1.15\,$mJ with the duration of $120\,$fs corresponds to the droplet fragmentation scenario when jets are formed at the droplet surface. The focused laser beam is tilted at $15^{\circ}$ relatively to the horizontal surface (Fig. ~\ref{fig:exp_setup}).
The focal length of the used lens is $17.5\,$cm. The irradiated droplet is positioned at the distance of $16.8$ cm from the lens, where the characteristic width of the laser beam (estimated as FWHM) is $200\,\mu$m. Thus the average laser pulse intensity in the target plane is $3\times10^{13}\,$W/cm$^2$.

The laser radiation polarization is filtered to be linear by situating a Glan--Taylor prism before the lens (Fig. ~\ref{fig:exp_setup}). The polarization in the target plane is adjusted by using quarter- and half-waveplates placed after the Glan--Taylor prism. Thus, using the quarter-waveplate, we are able to obtain circularly polarized radiation and using the half-waveplate we are able to rotate the radiation polarization plane. It is worth noting, that the droplet diameter is much smaller than characteristic beam size and the curvature radius of the radiation wave front ($\sim 8\,$mm) in the target plane. Thus, one can consider a droplet being uniformly irradiated by the laser with polarization defined by the used waveplate.

The droplet generator is synchronized with the laser, so that both are operating at the frequency of $4\,$Hz. Droplets are positioned with precision of $1\,\mu$m, and their response to a laser impact is reproduced well in series of experiments. Thus, the process of the droplet deformation with the following fragmentation can be studied using stroboscopic series of droplet snapshots which are taken at different time delays relatively to the Ti:Sapphire laser pulse. In our experiments such snapshot series are obtained using the method of instantaneous shadow photography. The instantaneity of the snapshots is ensured by a short duration of the back-light laser pulse, which is $\sim 30\,$ns. The moment, when the snapshot is taken, is controlled by varying time delay between the Ti:Sapphire laser pulse and the back-light laser pulse. The droplet response to the laser impact is recorded using the two CCD cameras positioned in one horizontal plane (Fig. ~\ref{fig:exp_setup}). The cameras are equipped with microscopes to take snapshots of relatively small droplets. The resolution of snapshots is $2.8\,\mu$m/pixel.

\section*{Experimental results}

The effect of laser pulse polarization on the fragmentation of liquid tin droplet was previously investigated in~\cite{de2021cylindrically}. It was shown that, using linearly polarized laser pulses, non-symmetric shock waves produced in the droplets lead to non-symmetric droplets deformation. In the work~\cite{de2021cylindrically} a fairly detailed analysis was carried out, but it mainly focuses on the initial stage of the droplet expansion, before the collapse of the cavity shell. Here we consider the late evolution of the droplet, when the collapse of the cavity shell results in jet effusion and the formation of secondary droplets. The late evolution of the droplet is determined by surface tension in contrast to the initial stage which is guided by shock and tensile waves.

\begin{figure}[t]
\centering\includegraphics[width=1.0\linewidth]{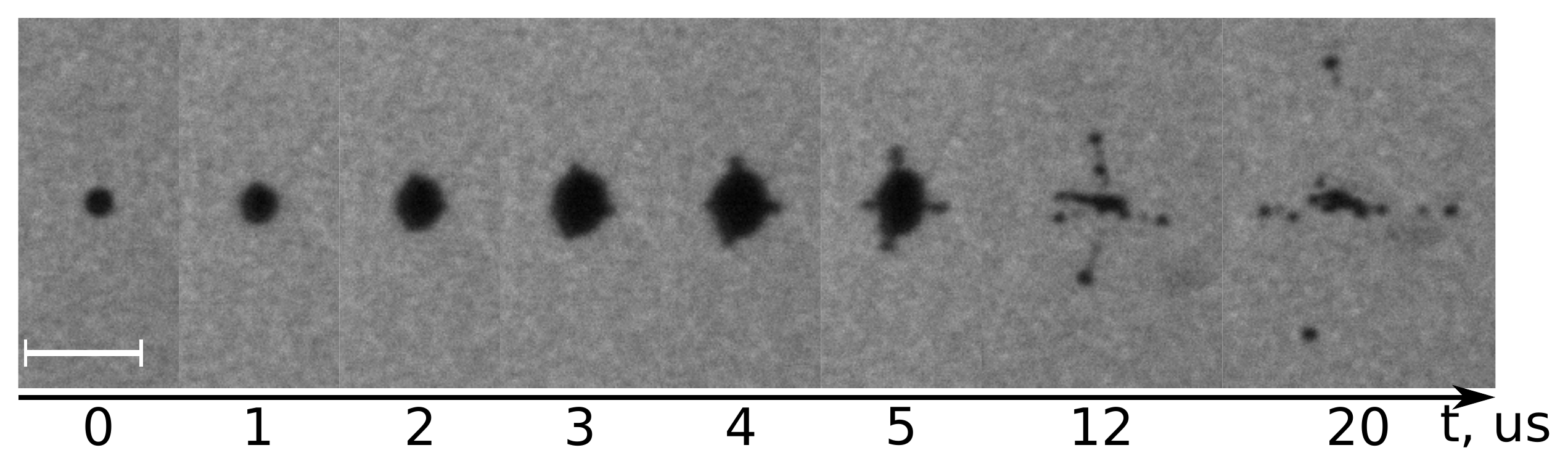} 
\caption{\label{fig:typical_scenario}
Droplet deformation and fragmentation in response to the linearly polarized laser pulse irradiation. The pulse arrives at time $t = 0\un{\mu s}$. The snapshots are shooted by the CCD2 camera, which provides a back view with angle $30^{\circ}$ to the horizontal projection of the laser beam axis as shown in Fig.~\ref{fig:exp_setup}. The length of the white scale line is $200\un{\mu m}$.}
\end{figure}

The typical scenario of the liquid metal droplet deformation and fragmentation observed with our experimental parameters in case of the linearly polarized laser radiation is presented in Fig.~\ref{fig:typical_scenario}.

Similar to the previously presented results~\cite{grigoryev2018expansion, krivokorytov:2018}, irradiation of a liquid metal droplet by a femtosecond laser pulse causes a rapid expansion with the spherical shape preserved. Since the mass of the metal ablated due to evaporation from the droplet surface is negligible, such expansion indicates the origin of a cavity expanding inside the droplet. It is worth noting that for all presented results (since the droplet diameter, the laser pulse energy, the spatial and temporal laser beam profiles are not varied) the specific energy of the laser radiation, which reached the droplet surface, amounts to $E_{OD}/m_D \approx 0.23\,$MJ/kg. According to the analysis presented in~\cite{grigoryev2018expansion}, the specific energy $E_{OD}/m_D$ should produce only a single bubble within the droplet, which is observed in Fig.~\ref{fig:typical_scenario}.

\begin{figure}[t]
\centering\includegraphics[width=1.0\linewidth]{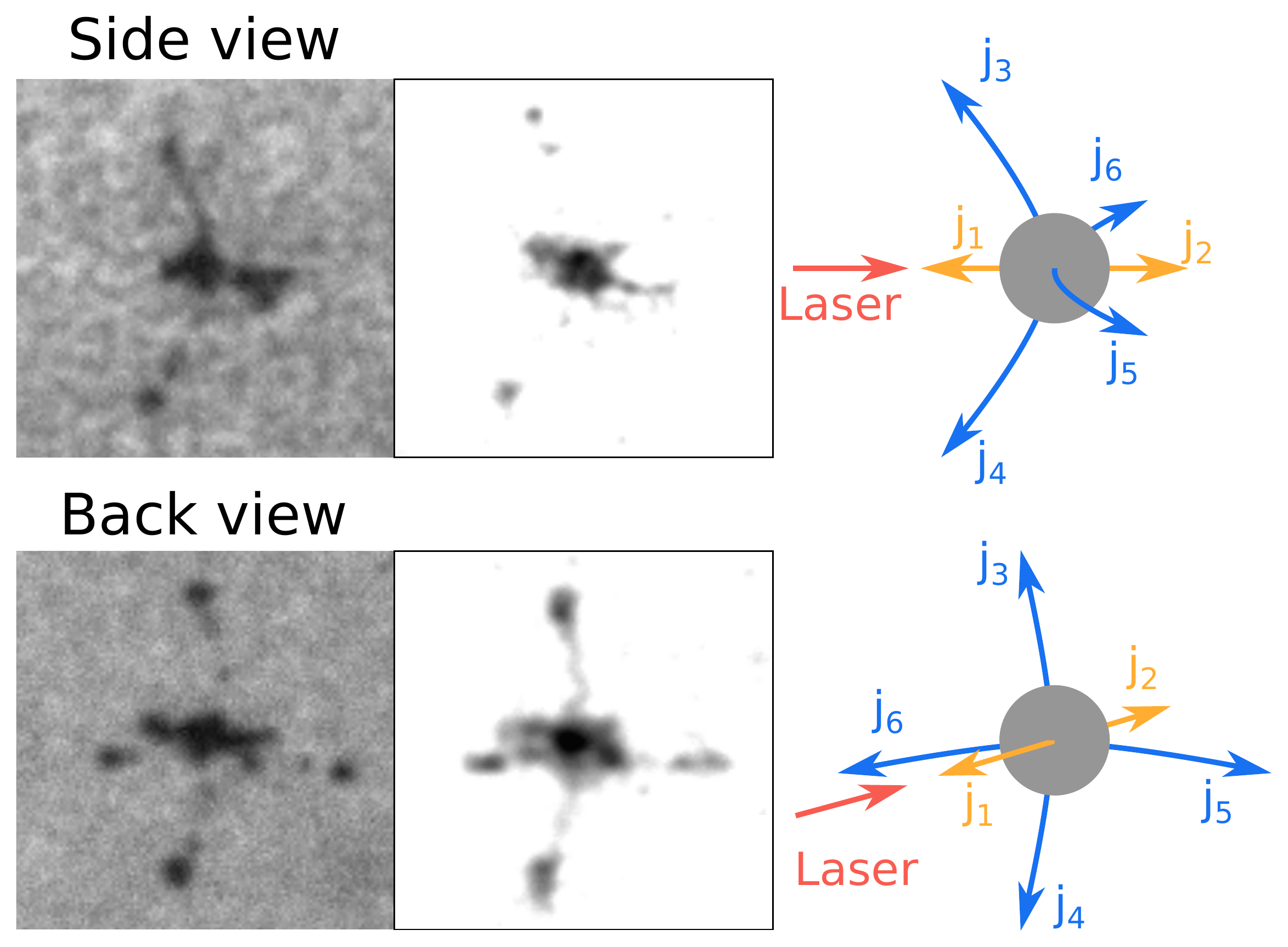} 
\caption{\label{fig:average_and_scheme}
Geometry of jets effused from the liquid tin droplet irradiated by a linearly polarized laser pulse. The top row represents the side views from CCD1 camera, the bottom row represents the back views from CCD2 camera with the angle of $30^{\circ}$ to the horizontal projection of the laser beam axis, see Fig.~\ref{fig:exp_setup}. The first column presents the corresponding snapshots of droplet shape at $12\un{\mu s}$ after the laser pulse. The second column presents the corresponding averaged image for the same time. The averaged image is obtained by overlaying snapshots with setting the grey scale proportional to probability that the material is present at a given position. Black pixels represent points where material is present in all snapshots, while white pixels indicate points where no material is observed in snapshots. The third column shows sketches of jets directions $\mathrm{j_1-j_6}$.}
\end{figure}

The observed shell of the bubble expands spherically during the first $3\un{\mu s}$ after the laser pulse irradiation.
Then that spherical symmetry of the shell is violated and humps become noticeable on the droplet surface as seen in Fig.~\ref{fig:typical_scenario}. At later times after the laser irradiation these humps evolve into jets. For linearly polarized laser light these jet directions are reproduced well in a series of experiments. To demonstrate this fact the image of droplet shape averaged over a hundred shots taken at the same time is presented in Fig.~\ref{fig:average_and_scheme}. The averaged image is obtained by overlaying snapshots of the droplet shape with setting the grey scale proportional to the probability that the material is present at a given position, see caption in Fig.~\ref{fig:average_and_scheme}. Similarity of the averaged image with the single shot indicates that the jets directions are reproduced well.

\begin{figure*}[t]
\centering\includegraphics[width=1.0\linewidth]{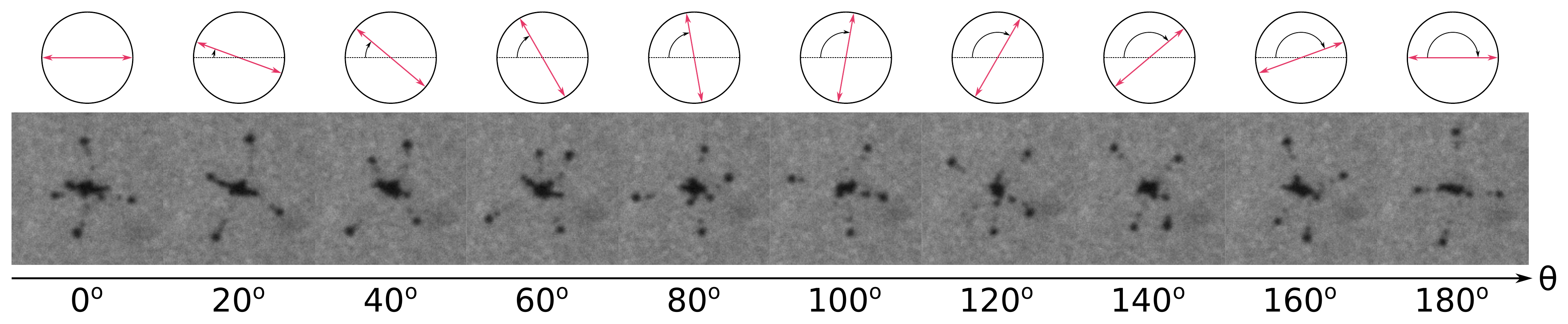} 
\caption{\label{fig:back_view_rotation}
		Snapshots of droplet shapes taken at $12\un{\mu s}$ after the linearly polarized laser pulse irradiation with the different polarization plane orientations. The top row of schemes with directions of $\theta$-axis represents the angle between the electric field vector and horizontal plane. The snapshots are obtained by the CCD2 camera, which shoots a back view with the angle $30^{\circ}$ to the horizontal projection of the laser beam axis as shown in Fig.~\ref{fig:exp_setup}.}
\end{figure*}

\begin{figure}[t]
\centering\includegraphics[width=1.0\linewidth]{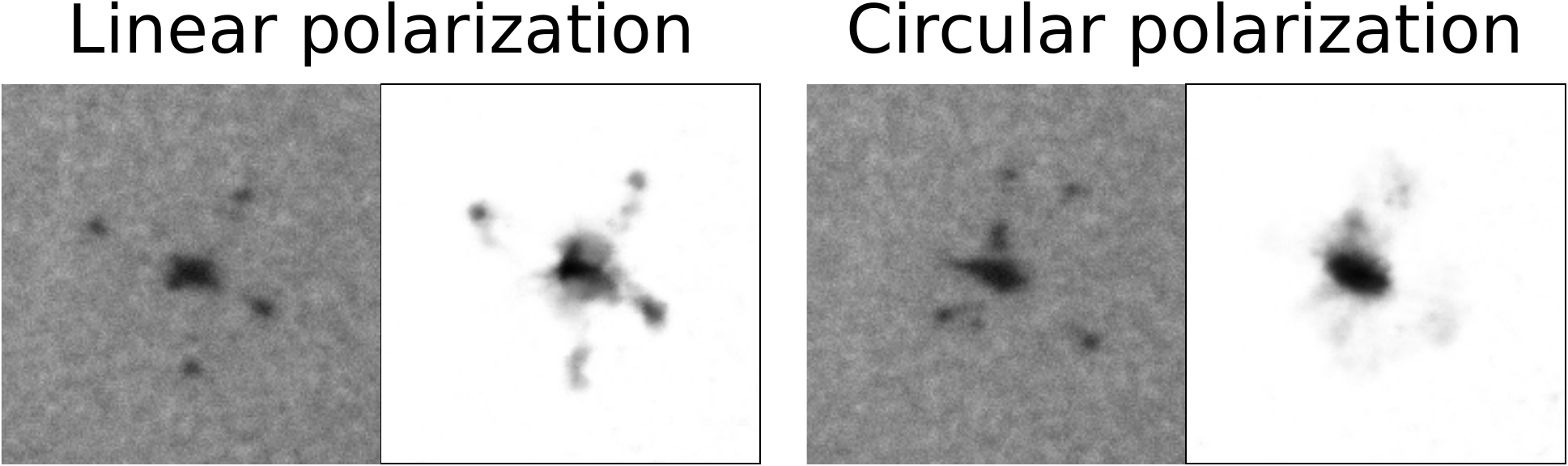} 
\caption {\label{fig:circular_polarization_res} 
Directions of laser-induced jets for linearly (left panel) and circularly polarized (right panel) laser pulses. Each panel contains two images: the left snapshot of droplet shape at $12\un{\mu s}$ after the pulse, and the right averaged image obtained by overlaying $450$ corresponding snapshots taken at the same time. All snapshots show the back views from CCD2 camera with angle $30^{\circ}$ to the horizontal projection of the laser beam axis, see Fig.~\ref{fig:exp_setup}.	}
\end{figure}

There is also a scheme of the directions of the induced jets presented in Fig.~\ref{fig:average_and_scheme}. 
It is convenient to denote each jet direction using labels $\mathrm{j_1-j_6}$. The observed jets can be divided into two groups: $\mathrm{j_1-j_2}$ is a pair of weak jets directed along the laser beam; $\mathrm{j_3-j_6}$ are the jets forming a cross-like structure in the plane perpendicular to the laser beam. Similar jets structure and directions were observed in our previous series of experiments~\cite{krivokorytov:2018}, but the mechanism of their formation was not discussed in details. The latter and the effect of laser radiation polarization on the jet formation mechanism is in focus of this study.

Influence of the linear polarized laser radiation is investigated for the orientation of a half-waveplate, see Fig.~\ref{fig:exp_setup}, i.e. by rotating the half-waveplate the polarization plane of the laser beam is also rotated. Such rotation does not change the integral quantity of absorbed laser energy, so the process of droplet expansion presented in Fig.~\ref{fig:typical_scenario} is identical for different orientations of the polarization plane. However, directions of the produced jets depend on the polarization plane orientation. The snapshots of fragmented droplets taken at the same time for different polarization plane orientation are presented in Fig.~\ref{fig:back_view_rotation}. One can notice that the directions of jets $\mathrm{j_3-j_6}$ are determined by orientation of the laser polarization plane, i.e. its electric field vector $\vec{E}$. In particular, the projections of jets $\mathrm{j_3-j_4}$ on the plane perpendicular to the laser axis are perpendicular to $\vec{E}$, while the projections of jets $\mathrm{j_5-j_6}$ on this plane are parallel to $\vec{E}$. Such connection between the jets directions and $\vec{E}$ leads to the rotation of jets $\mathrm{j_3-j_6}$ in the same direction as the polarization plane. It is also observed in the experiment that the velocities of jets $\mathrm{j_3-j_4}$ are larger than the velocities of jets $\mathrm{j_5-j_6}$ by a factor of $1.5$. Interestingly, the polarization plane orientation does not affect the directions and velocities of jets $\mathrm{j_1-j_2}$. It is worth noting that the observed correlation between the orientation of radiation polarization plane and the jets directions persists over a series of experiments.

We also conducted experiments with the a circular polarized laser pulse. The circular polarization is obtained by replacing the half-waveplate with a quarter-waveplate just positioned before the lens, see Fig.~\ref{fig:exp_setup}. Similar to results presented for linear polarization, this change affected only the directions of jets $\mathrm{j_3-j_6}$. The directions of these jets are now not anymore repeatable but vary between laser shots. This fact is demonstrated in Fig.~\ref{fig:circular_polarization_res} where the images from many experiments are averaged. The smearing (absence) of clear jets $\mathrm{j_3-j_6}$ in the averaged image indicates that each laser pulse produces jets along a seemingly random direction.

\section*{Molecular dynamics simulation}

Here we want to obtain an understanding of the physics resulting to the  the intriguing mechanism of jets formation. This may be achieved by continuum or atomistic simulations. The irradiated target experiences non-uniform heating which generates angle-dependent temperature and pressure distributions in the heated layer. Since after propagation of shock wave through a droplet the complex phenomena may happen, including formation of voids (cavitation) by the tensile stress and generation of jets with assistance of surface tension, a fluid dynamics approach would require {\em a priori} knowledge of the involved atomistic processes. Our previous approach~\cite{grigoryev2018expansion} was based on smoothed particle hydrodynamics method which allowed us to establish the droplet fragmentation mechanisms guided only by shock and tensile waves. However, a more realistic model that also includes surface tension deemed necessary to describe the present experiments. Thus, we now employ a molecular dynamics (MD) method as it covers in a realistic way all important physics once a valid inter-atomic potential is utilized.

The simulation results below are obtained with our in-house MD code MD-VD$^3$~\cite{Zhakhovskii:2005}. It allows to perform large-scale parallel simulations on a computing cluster keeping the optimal load balance between the nodes in tasks where atoms are non-uniformly distributed in a computational domain ~\cite{Egorova:2019}. However, similar to any MD problem, we are still restricted to a few billions of atoms with a micrometer-sized sample. Nevertheless, we have found that simulation of a droplet with the diameter larger than a hundred of nanometers provides results similar to experimental data, so that the phenomena observed in simulation can be scaled to experimental sizes.

\subsection{Inter-atomic potential for liquid tin}

Tin is a complex substance to model with MD because of its structural phase transitions in solid phase, so that both a simple pair potential and a more complex embedded atom model (EAM) potential can hardly reproduce tin properties. However, tin is fusible metal and the experiments deal only with liquid alloy, so that a EAM potential for liquid tin should be sufficient for simulation of liquid characteristics, including surface tension and liquid-vapor transition. The required potential must also reproduce the realistic shock Hugoniot of liquid tin.

The parameters of the potential are obtained using the stress-matching method developed in~\cite{Zhakhovskii:2009}, see also details on development technique in~\cite{Zhukhovitskii:2020}. They are adjusted to reproduce material behavior accurately in a given range of thermodynamic properties based on a fitting database. The construction of the fitting database is determined by the choice of atomic configurations, which correspond to states of the material at continuous cold compression. Thus, the database contains the equation of state for matter at absolute zero temperature, namely, the cold pressure $P(V)$ corresponding to uniform compression/tension. This choice for the fitting database ensures that the EAM-potential will describe the mechanical response at high compression/tension correctly. Since the thermal energy and thermal pressure are small compared to the potential energy of interaction between atoms and cold pressure in dense condensed phase (at relatively low temperature), the EAM-potentials obtained using the stress-matching method also give reasonable estimations for thermodynamic properties of the material at temperatures lower than critical.

Since we use an analytical form of EAM potential represented by the high-order rational functions, many local minima of target function exist in a multidimensional space of adjustable EAM parameters. Thus, the stress-matching method provides the several good EAM-potential candidates. To select the best candidate, all obtained EAM potentials are tested against available experimental data. The test results for the best candidate are discussed below.

\begin{figure}[t]
\centering\includegraphics[width=0.85\linewidth]{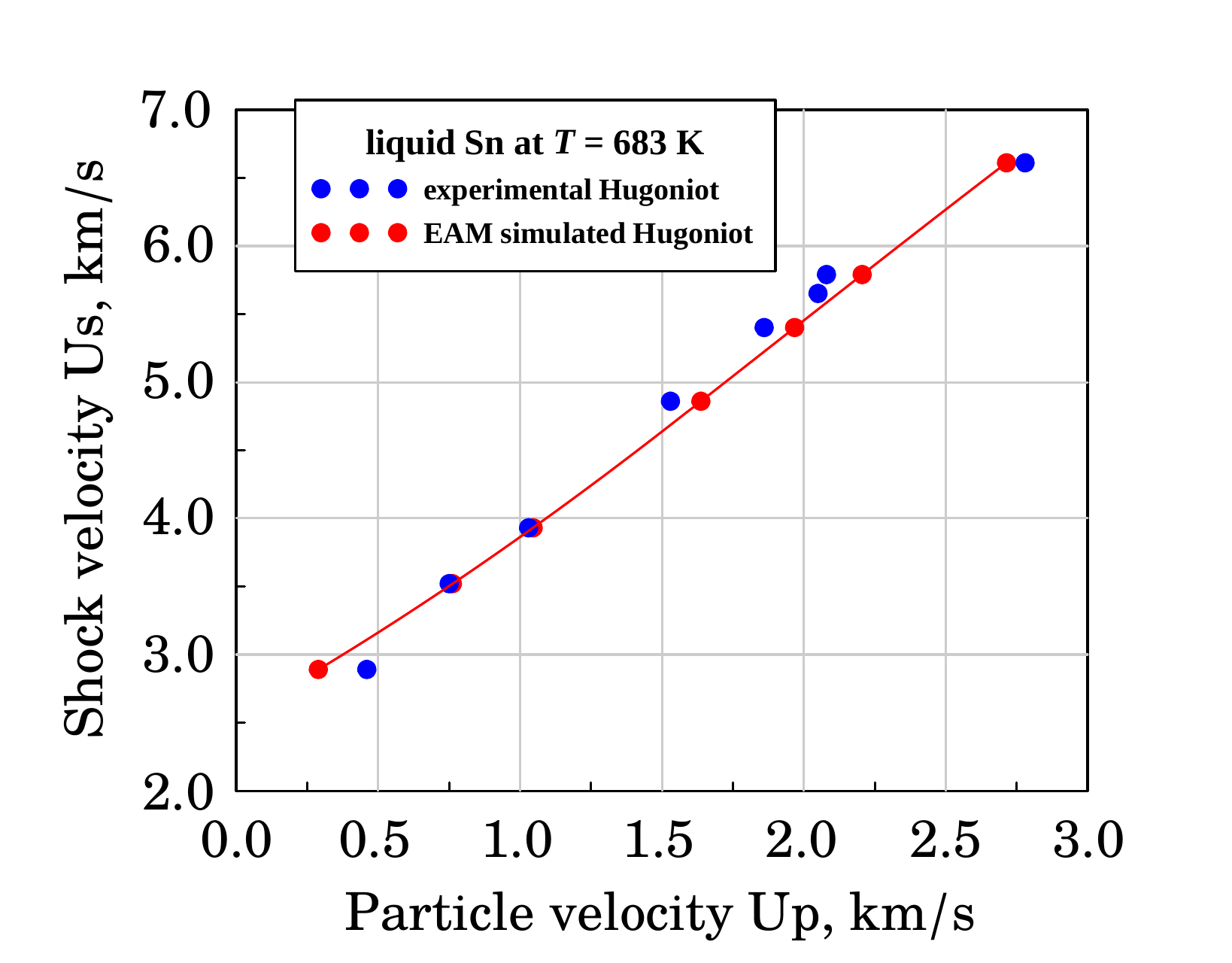} 
\caption {\label{fig:sn_hugoniot} 
Simulated and experimentally measured shock Hugoinot curves for liquid tin. The calculated points can be fitted by a linear function $\upsilon_{s} = c + a \upsilon_{p}$ with $c = 2366\un{m/s}$ and $a = 1.548$.}
\end{figure}

The first test of the several obtained EAM potentials is based on the comparison with experimental results of the shock wave propagation in the uncompressed molten tin at temperature $T=683\un{K}$. The Hugoniot curve obtained in MD simulation using the best potential  \cite{Zhakhovsky:EAM-project} can be approximated by equation:
\begin{equation}
\label{eq:hugoniot}
P(x)=\rho_{0}c^{2}\frac{x}{(1-a(1-x))^{2}},
\end{equation}
where $x=\rho_{0}/\rho$, the  density of uncompressed liquid tin $\rho_{0} = 7206\un{kg/m^3}$, the speed of sound $c = 2366\un{m/s}$ and the non-dimensional fitting parameter $a = 1.548$. Figure~\ref{fig:sn_hugoniot} shows comparison of calculated results with ones obtained in experiment~\cite{volkov1981investigation,trunin1995dynamic}. One can notice good agreement between the experiment and simulation.

The second test for the obtained EAM-potentials is to compare the experimentally measured surface tension of tin with the calculated one by its mechanical definition~\cite{rowlinson1982molecular}:
$
\sigma = \int \left(P_x(x) - P_y(x)\right)\,\mathrm{d}x,
$
where $x$-axis is perpendicular to an interface layer between liquid and vapor, $P_x(x)$ is the normal stress profile, and $P_y(x)$ is the tangential stress. The calculated surface tension with using the best potential is $\sigma_{MD} = 0.59\un{J/m^2}$, which agrees well with experimental surface tension $\sigma_{exp} \simeq 0.525\un{J/m^2}$ of liquid tin~\cite{kononenko1972surface, passerone1990influence, fima2010effect}.

\subsection{Absorption of linearly polarized light by spherical droplet}

\begin{figure*}[t]
\centering\includegraphics[width=1.0\linewidth]{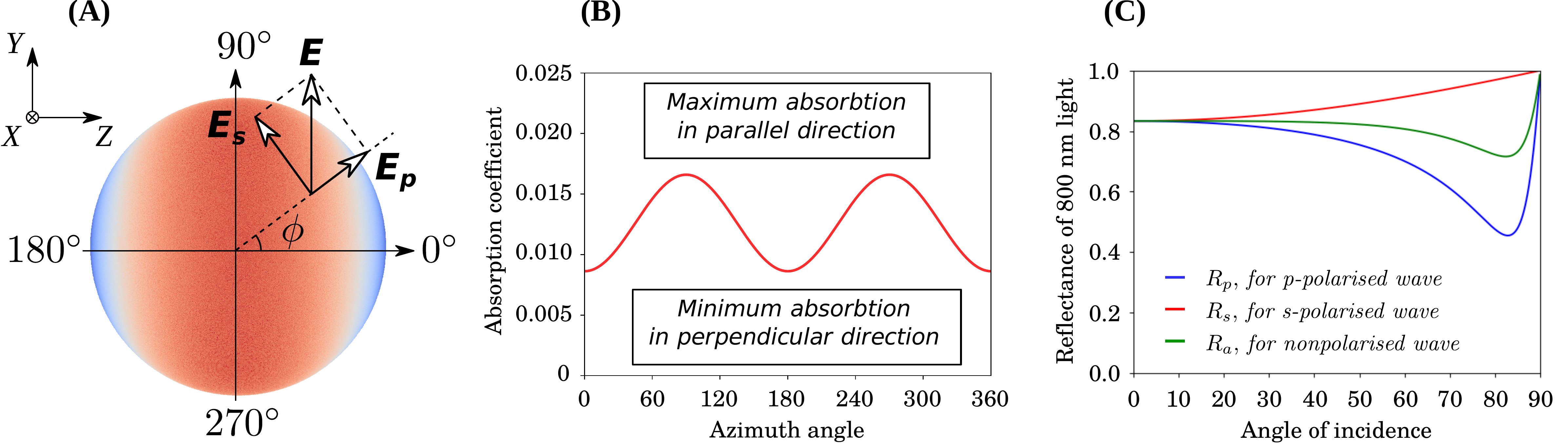}
\caption {\label{fig:reflect} 
(A) Sketch of frontal hemisphere heating by a linearly polarized light. Laser beam propagates along X-axis, the electric field vector $\vec{E}$ of radiation is parallel to Y-axis. $\vec{E}_s$ and $\vec{E}_p$ are s- and p-polarized components of the incident light. (B) Integral of the absorption coefficient over the incidence angle  $\theta$ as a function of the azimuth angle $\phi$. Absorption is the highest along the axis parallel to $\vec{E}$ and the lowest along the perpendicular axis. (C) Liquid tin reflection coefficients as functions of the light incidence angle for different polarizations according to Eqs.~(\ref{eq:Rs},\ref{eq:Rp}).}
\end{figure*}

Since both the laser pulse duration and the characteristic time of equilibrium attainment between electron and ion temperatures are much shorter than the sonic time $D/c_s\sim 10\un{ns}$ for the droplet with diameter of $D\sim 30\un{\mu m}$, the laser energy absorption can be considered as an instantaneous process. Therefore, in our simulations the laser pulse absorption is simulated by setting an initial temperature distribution in the heated layer on the irradiated frontal hemisphere of a droplet.

The linearly polarized laser pulse is supposed to provide the expected temperature distribution within a droplet after electron-ion equilibration (during which the heated layer is formed):
\begin{equation}
\label{eq:temperature}
T(r, \theta, \phi) = T_{m} \alpha(\theta, \phi)  \exp[-(r - R)^2/\delta^2_h],
\end{equation}
where $\theta$ is an angle of light incidence (between the normal to the sphere surface in a given point and the laser beam axis), $\phi$ is an azimuth angle between the light incidence plane and the polarization plane (Fig.~\ref{fig:reflect}(A)), $\alpha (\theta, \phi)$ is a local absorption coefficient, $T_m$ is a temperature parameter on the irradiated pole of a droplet.
The exponent term in Eq.~(\ref{eq:temperature}) describes the exponential decrease of temperature with distance from the droplet's surface. Thus, $r$ is the coordinate along the droplet radius, $R$ is the droplet radius, and $\delta_h$ is the characteristic depth of the heated layer.

As it is mentioned above in Introduction, MD simulation of a droplet with diameter $\sim 30\un{\mu m}$ cannot be performed directly --- such a big sample would contain a huge number of atoms of about $10^{15}$. Instead, a liquid tin droplet with diameter of $160\un{nm}$ containing a reasonable number of atoms of about $10^8$ is simulated by MD method, and then the proper scaling of simulation dimensions to experimental sizes is applied to compare the corresponding results.

\begin{figure*}[t]
\centering\includegraphics[width=1.0\linewidth]{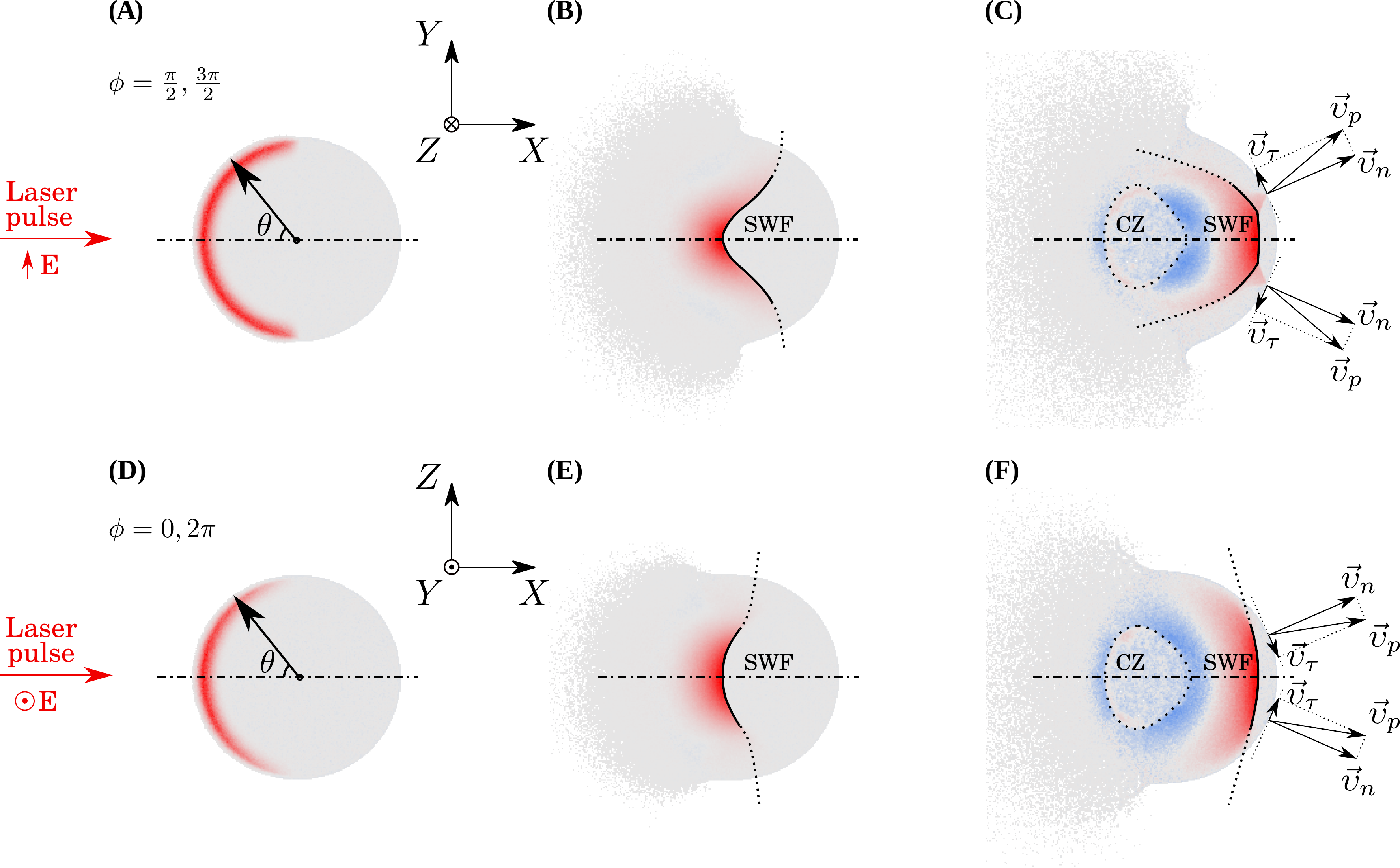}
\caption {\label{fig:swf} 
Cross-sections of pressure distribution in shock wave propagating along the X-axis from the irradiated frontal surface of the droplet to its rear surface ($\theta$ is the incidence angle, $\phi$ is the azimuth angle). The top maps show the XY-plane parallel to the vector of electromagnetic field $\vec{E}$, the bottom maps show the XZ-plane perpendicular to $\vec{E}$. Red color corresponds to compression, blue color --- to tension. ``SWF'' stands for the shock wave front, and ``CZ'' for the cavity zone. The rear surface velocity is represented by vector $\vec{\upsilon}_p$, but $\vec{\upsilon}_n$ is the velocity component normal to the rear surface, and $\vec{\upsilon}_{\tau}$ is the tangential velocity. The dash-dotted line indicates the symmetry plane.}
\end{figure*}

Such similarity between deformation of droplets with different diameters in the experimental conditions corresponding to ours was studied in our previous work~\cite{grigoryev2018expansion}. According to the study, the droplet deformation and fragmentation in the wide range of droplet sizes is completely defined by the ratio between the laser energy absorbed in a droplet and the droplet mass. Wherein the characteristic thickness of the heated layer does not influence the droplet dynamic at later times. However, these conclusions were made for the case when the thickness of the heated layer (and so the initial width of the laser-induced shock-wave front) was much smaller than the droplet radius. When the droplet diameter equals to $160\,$nm, the skin layer (and the depth of the heated layer) is comparable to the droplet radius. Therefore, simulations of the droplet fragmentation using MD method, which can be properly scaled to the results observed in experiment, should be performed with the deliberately decreased depth of the heated layer, while keeping the ratio of absorbed laser energy to the droplet mass the same as in the experiment. The adjusted depth of the heated layer should be much smaller than the droplet radius, but much larger than the distance between atoms. In our simulations $\delta_h = 8\un{nm}$ is used.

It is worth noting, that the droplet deformation presented in~\cite{grigoryev2018expansion} were obtained without taking into account surface tension. And as it was discussed in~\cite{krivokorytov:2018}, surface tension should not be neglected for the presented regime of the laser-induced droplet fragmentation. Therefore, we can expect only qualitative agreement between MD model and experimental results at early times after the laser pulse impact. Kinetic energy of the droplet expansion decreases with time, which leads to the increase of the surface tension effect at later times, which, in turn, may lead to the discrepancy between MD model and experimental results.

Absorption coefficient  $\alpha = \alpha (\theta, \phi)$ of linearly polarized light is a function of the incidence angle $\theta$ and the azimuth angle $\phi$. One can decompose the electric field vector $\vec{E}$ of linearly polarized light into two orthogonal linearly polarized components. If one consider such decomposition relatively to the surface the light incidence, these two components would coincide with s- and p-polarized radiation. For the spherical surface and the linearly polarized incident light with $\vec{E}$, the s-polarization electric field component is $\vec{E_s} = \vec{E} \cos \phi$ and the p-polarized electric field component is $\vec{E_p} = \vec{E} \sin \phi$ as illustrated in Fig.~\ref{fig:reflect}(A). According to this decomposition the absorption coefficient $\alpha (\theta, \phi)$ can be written as:
\begin{equation}
\label{eq:absorption}
\alpha(\theta, \phi) = [(1 - R_s)\cos^2\phi + (1 - R_p)\sin^2\phi] \cos\theta,
\end{equation}
where $R_s$ and $R_p$ are the reflection coefficients of s- and p-polarized light, respectively. Here, the first term in the squared brackets represents the absorption of s-polarized component and the second is the absorption of p-polarized component. The coefficient $\cos \theta$ in Eq.~(\ref{eq:absorption}) occurs due to the decrease of the irradiated surface area when approaching the droplet's equator. According to the generalized Fresnel formulas~\cite{palik1998handbook}  the reflection coefficients $R_{s}$ and $R_{p}$ shown in  Fig.~\ref{fig:reflect}(C) can be written as
\begin{equation}
\label{eq:Rs}
R_{s} = \frac{(a - \cos \theta)^2 + b^2}{(a + \cos \theta)^2 + b^2},
\end{equation}
\begin{equation}
\label{eq:Rp}
R_{p} = R_{s} \frac{(a - \sin \theta \tan \theta)^2 + b^2}{(a + \sin \theta \tan \theta)^2 + b^2},
\end{equation}
\begin{eqnarray}
\label{eq:a2}
a^{2} = \frac{1}{2}\left ( \left( n^{2} - k^{2} - \sin^{2}\theta \right )^{2} + 4n^{2}k^{2} \right )^{\frac{1}{2}} & & \nonumber \\
+ \frac{1}{2} \left ( n^{2} - k^{2} - \sin^{2}\theta \right ), & &
\end{eqnarray}
\begin{eqnarray}
\label{eq:b2}
b^{2} = \frac{1}{2}\left( \left ( n^{2} - k^{2} - \sin^{2}\theta \right )^{2} + 4n^{2}k^{2} \right )^{\frac{1}{2}} & & \nonumber \\
- \frac{1}{2} \left ( n^{2} - k^{2} - \sin^{2}\theta \right ), & &
\end{eqnarray}
where $n$ and $k$ are the real and imaginary parts of the material refractive index. In our simulations $n=2.96$ and $k=7.44$ are used which were experimentally measured for radiation with the wavelength $\lambda = 800\,$nm in~\cite{petrakian1980optical}.

It should be emphasized that since the absorption coefficient is a function of the incidence and azimuth angles it varies with positions on the spherical droplet surface. Therefore, even for uniform irradiation of the droplet's surface by the linearly polarized laser pulse the absorbed energy distribution on the droplet surface (and so the surface temperature) will not be uniform.

\begin{figure*}[t]
\centering\includegraphics[width=1.0\linewidth]{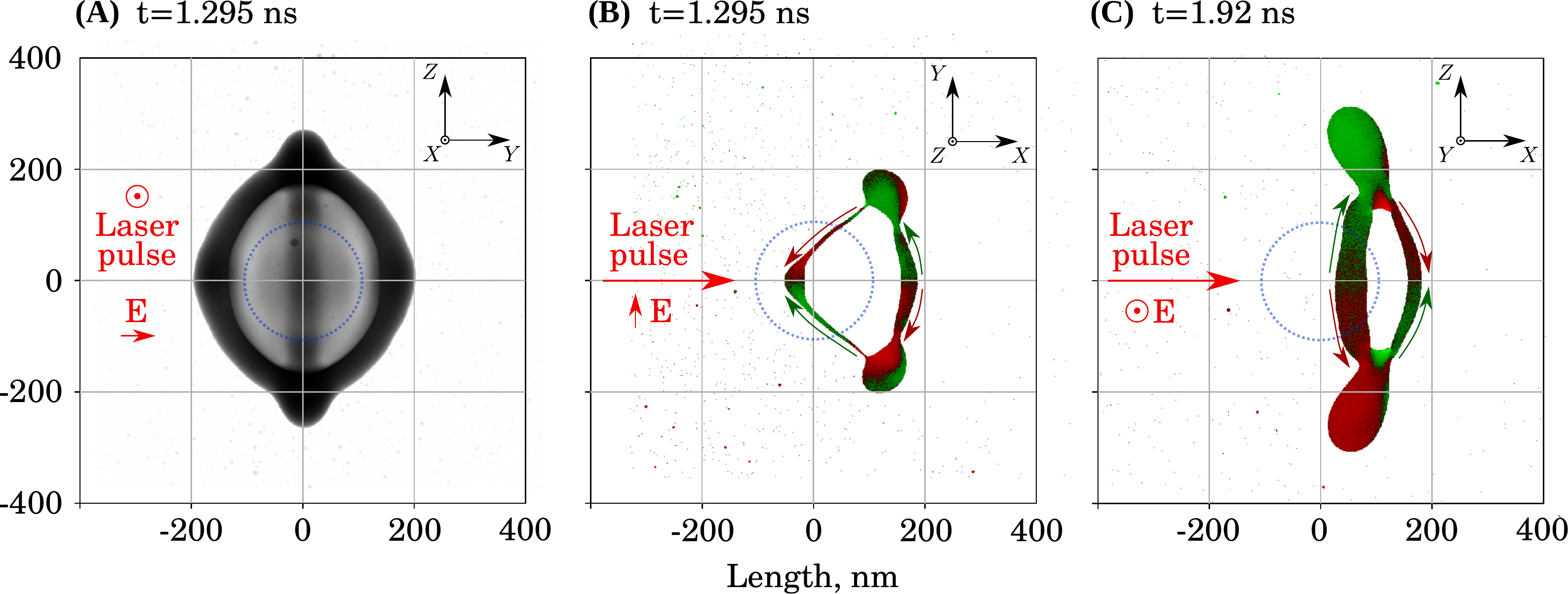}
\caption {\label{fig:maps1} 
(A) Droplet shape after the cavity shell formation. The flow on the frontal part of the cavity shell accumulates liquid along the direction perpendicular to vector $\vec{E}$. As a result, a bead is formed along the axis Z. (B,C) Flow field in the two perpendicular planes (green color corresponds to positive velocity, red color --- to negative velocity). In the XY-plane, a converging flow is formed on the frontal surface of the shell, and a diverging flow is formed on the rear surface (B). In the XZ-plane, the flow is inverted: the front surface is elongated, and a converging flow is formed on the rear surface (C).}
\end{figure*}

In order to demonstrate variations of the absorbed energy distribution the  integral of the absorption coefficient ${\overline\alpha}_{\phi}=\int\alpha(\theta,\phi)\mathrm{d}\theta$ over the incidence angle $\theta$ is presented as a function of azimuth angle $\phi$ in Fig.~\ref{fig:reflect}(B). One can see that the absorption of linearly polarized laser radiation is maximal along the axis parallel to the polarization plane, and the absorption is minimal along the axis  perpendicular to the polarization plane. The ratio between the highest and the lowest absorption coefficients ${\overline\alpha}_{\phi}$ is approximately $2$.

\subsection{Simulation results}

Here the molecular dynamics simulation of the droplet fragmentation is carried out for the low intensity regime of irradiation. We want to emphasis, that we can't provide a quantitative comparison as the size of the simulated droplet differs from the experimental droplet size by two orders of magnitude. Previously we have shown that droplets of different sizes expand in a similar way if the specific absorbed energy, i.e. thus normalized by the droplet mass, remains the same~\cite{grigoryev2018expansion}. As shown below, we again obtain quantitative similar results when the specific absorbed energy per droplet mass is $0.4\un{kJ/g}$, which is comparable to the experimental energy of $0.23\un{kJ/g}$. Applying this higher specific energy in MD is motivated by the fact, that the specific surface energy (per droplet mass) of large experimental droplets is considerably lower than that of small droplets simulated by MD method. 

The physical picture of the expansion and fragmentation of a liquid tin droplet irradiated by a polarized ultrashort laser pulse shares some similarities with the fragmentation after irradiation with unpolarized light~\cite{grigoryev2018expansion}. However, the specific shock loading caused by irradiation with linearly polarized light results into distinct expansion features observed in the experiment. Figure~\ref{fig:swf} shows the simulated pressure maps at different times during propagation of a shock pulse from the frontal irradiated surface layer to the rear surface of droplet. The red and blue colors indicate the location of compression (positive pressure) and tension (negative pressure), respectively. The top Figs.~\ref{fig:swf}(A-C) demonstrate the motion of  shock pulse in the droplet slice along the XY-plane, i.e. perpendicular to the vector of electromagnetic field $\vec{E}$. The bottom Figs~\ref{fig:swf}(D-F) show pressure distributions in XZ-plane perpendicular to the vector of electromagnetic field $\vec{E}$.

It should be emphasized here, that below we consider only the angle-dependent components of pressure gradients responsible for non-spherical expansion of droplet and jet configuration. The radial expansion is produced by radial-dependent forces discussed before in our work \cite{grigoryev2018expansion}. Heating of a thin frontal surface layer after absorption of the laser pulse forms a nonuniform pressure field having a maximum gradient along the incidence angle $\theta$ in the XZ-plane, see Fig.~\ref{fig:swf}(A,D). The pressure gradient in the $\theta$ direction is significantly smaller in the XY-plane, except the region close to the equatorial plane with $x=0$, see Fig.~\ref{fig:swf}(A). Therefore, such an asymmetric heating results in a nearly hemispherical ablated plume in the XY-plane (Fig.~\ref{fig:swf}(B,C)) and a more backward-directed plume in the XZ-plane (Fig.~\ref{fig:swf}(E,F)). Similar distributions of plume material were also observed in experiments~\cite{de2021cylindrically}. This asymmetric heating generates also a diverging liquid flow with nearly $\phi=0$ on the frontal surface from the droplet pole at $\theta=0^{\circ}$ to the equatorial zone at $\theta=90^{\circ}$. And another converging flow arises on the frontal surface along the azimuthal direction towards the XZ-plane.

\begin{figure*}[t]
\centering\includegraphics[width=1.0\linewidth]{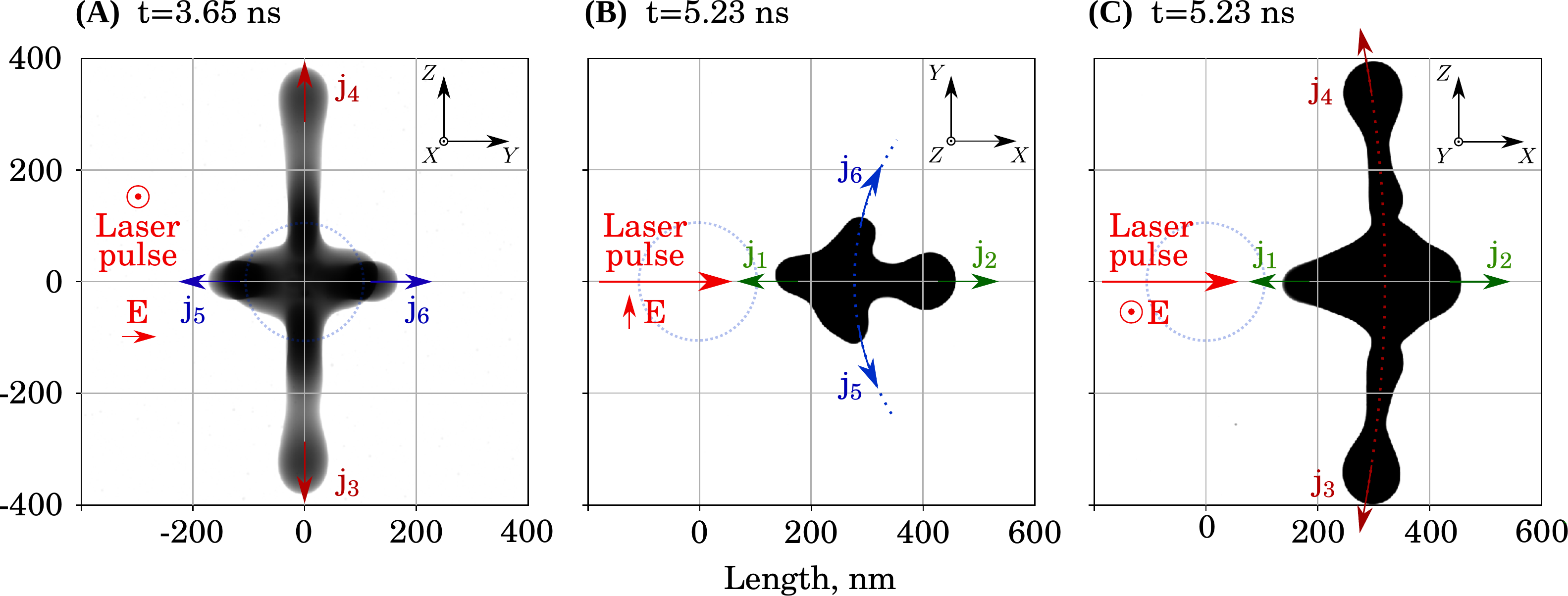}
\caption {\label{fig:maps2}
Evolution of the droplet shape after collapse of the cavity shell. (A) Cross-like structure of jets $\mathrm{j_3-j_6}$ is formed in the plane perpendicular to the laser pulse. (B) Short low-speed jets $\mathrm{j_5-j_6}$ formed from the rearside shell are directed parallel to electromagnetic field vector $\vec{E}$. In addition two tiny jets $\mathrm{j_1-j_2}$ directed along the laser pulse are formed at a later time. (C) Long high-speed jets $\mathrm{j_3-j_4}$ formed from the frontal side are directed perpendicular to $\vec{E}$.}
\end{figure*}

Figure~\ref{fig:maps1} shows the transitional shape of the droplet after formation of a central cavity. The cavity shell has the thinnest frontal part in the area of maximum shock loading, which is caused by the azimuthal flow from this region as illustrated by Fig.~\ref{fig:maps1}(B). Eventually, it leads to rupture of the shell. It is seen that the liquid begins to accumulate on the frontal part of cavity shell and stretch along Z-axis. This flow contributes to shell extension along the direction perpendicular to the vector $\vec{E}$ as shown in Fig.~\ref{fig:maps1}(C). Thus, the pair of jets $\mathrm{j_3-j_4}$ observed in our experiments is formed from the frontal part of the shell as a result of asymmetric shock loading after absorption of a linearly polarized laser pulse. 

Figure~\ref{fig:maps2} shows the final stage of droplet expansion. Here, the formation of jets $\mathrm{j_3-j_4}$ is clearly seen along the direction perpendicular to the vector $\vec{E}$. The expansion of jets $\mathrm{j_3-j_4}$ is not exactly aligned with the axis, but those ends are bent towards the laser pulse as seen in Fig.~\ref{fig:maps2}(C). Thus, the frontal part of the cavity shell is everted under the asymmetric pressure produced by ablating frontal material. A similar picture is observed in the experiment, see Fig.~\ref{fig:average_and_scheme}.

Figures~\ref{fig:swf}(B,E) demonstrate the perpendicular cross-sections of pressure distribution in the shock wave front (SWF) focusing near the droplet center. It should be noted, that the direction of material flow induced by reflection of oblique shock front from a free surface depends on a shock incidence angle. In general, the curved shock front may cause the various angular material motions in a subsurface layer, including flows converging and diverging with angle, after arrival to a spherical surface of droplet.  

Indeed, the SWF in the XY-plane has a steep curvature, as seen in Fig.~\ref{fig:swf}(B), while in the XZ-plane it is almost flat, see Fig.~\ref{fig:swf}(E). Reflection of such a curved SWF from the rear surface of droplet induces a diverging flow in the XY-plane, since the curvature radius of the SWF in this plane is smaller than the droplet radius, as shown by $\vec{\upsilon}_{\tau}$ in Fig.~\ref{fig:swf}(C). In contrast, the flow in XZ-plane is converging as seen in $\vec{\upsilon}_{\tau}$ in Fig.~\ref{fig:swf}(F). Thus, reflection of the almost flat SWF from the droplet rear surface accumulates liquid there along the azimuth angle $\phi=3 \pi / 2$, i.e. in the directions of the Y-axis.

Figure~\ref{fig:swf}(C,F) shows the local velocity $\vec{\upsilon}_p$ after reflection of the shock wave from the rear surface of the droplet. The normal velocity $\vec{\upsilon}_n$ determines the expansion of the cavity shell in the radial direction. The tangent velocity $\vec{\upsilon}_{\tau}$ induces a flow along the rear surface of the droplet. The velocity field of the cavity shell reveals an inverse flow pattern in the perpendicular planes, shown in Fig.~\ref{fig:maps1}(B,C). The velocity field on the rear surface leads to the droplet elongation along the direction parallel to the vector $\vec{E}$. It should be noted that the velocity field on the rear surface is established later than on the frontal surface of droplet.  Because of this the map in Fig.~\ref{fig:maps1}(C) is shown at a later time of $1.92\un{ns}$, after the rupture of the frontal side of cavity shell. 

Figure~\ref{fig:maps2} reveals the late evolution of the simulated droplet shape. Jets $\mathrm{j_5-j_6}$ along the Y-direction are formed due to the flow induced by reflection of the shock wave from the rear surface of droplet. It is also worth noting, that the expansion of jets $\mathrm{j_5-j_6}$ occurs not exactly along the direction of Y-axis, but slightly curved towards the laser pulse. Overall, by comparing Fig.~\ref{fig:maps2} with the experimental results represented in Fig.~\ref{fig:average_and_scheme}  we find a good correlation in the occurrence and directions of filamental jets.

Yet, in MD simulations, where the droplet size is several orders of magnitude smaller than in our experiments, expansion of jets $\mathrm{j_5-j_6}$ does not lead to their fragmentation. This is fully understandable, as the surface tension dominates over inertial expansion during the later stage of fragmentation. It stops the jet elongation and prevents development of the Plateau-Rayleigh instability. An increase of laser fluence would increase the expansion rate in simulation. However, it would also make the shell thinner, and rupture of the rear part of shell would occur before formation of jets. Thus, our simulations demonstrate the possibility of jet formation in the direction parallel to vector $\vec{E}$. On the other hand, MD simulation of jet fragmentation requires a significant increase in the spatial scale, which may only be realized with the continuum hydrodynamic methods, but those need to take into account the surface tension.

Further evolution of jets $\mathrm{j_3-j_6}$ in experimental conditions leads to their fragmentation via the Plateau-Rayleigh instability, but at small MD scale the surface tension prevents the required elongation of those jets and induce a converging flow near their base. As a result two tiny jets $\mathrm{j_1-j_2}$ are formed along the laser pulse at a later time, see the corresponding green arrows shown in Fig.~\ref{fig:maps2}(B,C). We also observe formation of jets $\mathrm{j_1-j_2}$ in our experiments, which are presented in Fig.~\ref{fig:average_and_scheme}.

\section*{Conclusions} 

Angle-dependent jet effusions from the liquid tin droplets irradiated by linearly polarized ultrashort laser pulses are observed in shadowgraph snapshots obtained in our experiments. To reveal the underlying mechanism of jet formation the complex sequence of processes is simulated using the molecular dynamics method. As we found the main features of droplet fragmentation have much in common with earlier reported ones produced with unpolarized laser pulses \cite{grigoryev2018expansion}. Here again, the absorption of light in a thin frontal surface layer generates a strong pressure pulse propagating through a droplet, which results to rupture of the liquid, i.e. cavitation, near the droplet center. Expansion of the droplet shell with an interior cavity inside leads to the shell breakup into smaller fragments distributed uniformly around the axis of laser propagation. In the present experiments using linearly polarized light, the absorption coefficient depends now on the orientation of electromagnetic field, which results in angle-dependent expansion and fragmentation of the droplet at a later time.

Recently we demonstrated experimentally \cite{grigoryev2018expansion} that the laser beam with circularly polarized or unpolarized light results in formation of jets during the droplet expansion, while their number and directions are stochastic. In this work with the use of linearly polarized light, we discover the formation of cruciform jets ejected away in roughly perpendicular directions. Moreover, it is demonstrated by rotating the polarization vector that the jet directions are strictly determined by the electric field vector of laser light. There is a pair of high-speed jets oriented across the electric field vector $\vec{E}$ and another pair of low-speed jets oriented along $\vec{E}$. We also observed a pair of tiny jets directed along the laser axis, which are formed later on.

To understand the dynamics leading to the experimentally observed, we performed molecular dynamics simulation of droplet expansion induced by angle-dependent heating. The almost isochoric heating of the thin surface layer depends on the angle between a polarization vector and the surface normal of the droplet, which result in a cross flow pattern on the frontal (irradiated) and rear surfaces. We demonstrated that the flow velocity field within the frontal layer of droplet, which is formed as a result of heating, leads to the formation of two opposite jets oriented across the electric field vector $\vec{E}$. Conversely, reflection of the shock wave from the rear surface accelerates material in such a way that the impulsively generated flow field on the rear side generates two opposite jets oriented along the polarization vector, $\vec{E}$. The MD simulation also revealed, that the pair of frontal jets is faster than the pair of rear side jets; again in agreement with the  experiments. Surface tension leads to the formation of two tiny jets along the laser beam late on the droplet fragmentation.

\acknowledgments{} 
The experimental part of this work was supported by the Russian Foundation for Basic Research under Grant No. 20-21-00143-ROSATOM. The theoretical part was supported by the Russian Science Foundation, Grant No. 19-19-00697.


\begin{thebibliography}{32}%
\makeatletter
\providecommand \@ifxundefined [1]{%
 \@ifx{#1\undefined}
}%
\providecommand \@ifnum [1]{%
 \ifnum #1\expandafter \@firstoftwo
 \else \expandafter \@secondoftwo
 \fi
}%
\providecommand \@ifx [1]{%
 \ifx #1\expandafter \@firstoftwo
 \else \expandafter \@secondoftwo
 \fi
}%
\providecommand \natexlab [1]{#1}%
\providecommand \enquote  [1]{``#1''}%
\providecommand \bibnamefont  [1]{#1}%
\providecommand \bibfnamefont [1]{#1}%
\providecommand \citenamefont [1]{#1}%
\providecommand \href@noop [0]{\@secondoftwo}%
\providecommand \href [0]{\begingroup \@sanitize@url \@href}%
\providecommand \@href[1]{\@@startlink{#1}\@@href}%
\providecommand \@@href[1]{\endgroup#1\@@endlink}%
\providecommand \@sanitize@url [0]{\catcode `\\12\catcode `\$12\catcode
  `\&12\catcode `\#12\catcode `\^12\catcode `\_12\catcode `\%12\relax}%
\providecommand \@@startlink[1]{}%
\providecommand \@@endlink[0]{}%
\providecommand \url  [0]{\begingroup\@sanitize@url \@url }%
\providecommand \@url [1]{\endgroup\@href {#1}{\urlprefix }}%
\providecommand \urlprefix  [0]{URL }%
\providecommand \Eprint [0]{\href }%
\providecommand \doibase [0]{http://dx.doi.org/}%
\providecommand \selectlanguage [0]{\@gobble}%
\providecommand \bibinfo  [0]{\@secondoftwo}%
\providecommand \bibfield  [0]{\@secondoftwo}%
\providecommand \translation [1]{[#1]}%
\providecommand \BibitemOpen [0]{}%
\providecommand \bibitemStop [0]{}%
\providecommand \bibitemNoStop [0]{.\EOS\space}%
\providecommand \EOS [0]{\spacefactor3000\relax}%
\providecommand \BibitemShut  [1]{\csname bibitem#1\endcsname}%
\let\auto@bib@innerbib\@empty
%</preamble>
\bibitem [{\citenamefont {Hirano}\ \emph {et~al.}(2001)\citenamefont {Hirano},
  \citenamefont {Komatsu}, \citenamefont {Saeki}, \citenamefont {Uenohara},
  \citenamefont {Takahashi}, \citenamefont {Takayama},\ and\ \citenamefont
  {Yoshimoto}}]{hirano2001enhancement}%
  \BibitemOpen
  \bibfield  {author} {\bibinfo {author} {\bibfnamefont {T.}~\bibnamefont
  {Hirano}}, \bibinfo {author} {\bibfnamefont {M.}~\bibnamefont {Komatsu}},
  \bibinfo {author} {\bibfnamefont {T.}~\bibnamefont {Saeki}}, \bibinfo
  {author} {\bibfnamefont {H.}~\bibnamefont {Uenohara}}, \bibinfo {author}
  {\bibfnamefont {A.}~\bibnamefont {Takahashi}}, \bibinfo {author}
  {\bibfnamefont {K.}~\bibnamefont {Takayama}}, \ and\ \bibinfo {author}
  {\bibfnamefont {T.}~\bibnamefont {Yoshimoto}},\ }\href {\doibase
  10.1002/lsm.1129} {\bibfield  {journal} {\bibinfo  {journal} {Lasers in
  Surgery and Medicine: The Official Journal of the American Society for Laser
  Medicine and Surgery}\ }\textbf {\bibinfo {volume} {29}},\ \bibinfo {pages}
  {360} (\bibinfo {year} {2001})}\BibitemShut {NoStop}%
\bibitem [{\citenamefont {Nakagawa}\ \emph {et~al.}(2002)\citenamefont
  {Nakagawa}, \citenamefont {Hirano}, \citenamefont {Komatsu}, \citenamefont
  {Sato}, \citenamefont {Uenohara}, \citenamefont {Ohyama}, \citenamefont
  {Kusaka}, \citenamefont {Shirane}, \citenamefont {Takayama},\ and\
  \citenamefont {Yoshimoto}}]{nakagawa2002holmium}%
  \BibitemOpen
  \bibfield  {author} {\bibinfo {author} {\bibfnamefont {A.}~\bibnamefont
  {Nakagawa}}, \bibinfo {author} {\bibfnamefont {T.}~\bibnamefont {Hirano}},
  \bibinfo {author} {\bibfnamefont {M.}~\bibnamefont {Komatsu}}, \bibinfo
  {author} {\bibfnamefont {M.}~\bibnamefont {Sato}}, \bibinfo {author}
  {\bibfnamefont {H.}~\bibnamefont {Uenohara}}, \bibinfo {author}
  {\bibfnamefont {H.}~\bibnamefont {Ohyama}}, \bibinfo {author} {\bibfnamefont
  {Y.}~\bibnamefont {Kusaka}}, \bibinfo {author} {\bibfnamefont
  {R.}~\bibnamefont {Shirane}}, \bibinfo {author} {\bibfnamefont
  {K.}~\bibnamefont {Takayama}}, \ and\ \bibinfo {author} {\bibfnamefont
  {T.}~\bibnamefont {Yoshimoto}},\ }\href {\doibase 10.1002/lsm.10055}
  {\bibfield  {journal} {\bibinfo  {journal} {Lasers in surgery and medicine}\
  }\textbf {\bibinfo {volume} {31}},\ \bibinfo {pages} {129} (\bibinfo {year}
  {2002})}\BibitemShut {NoStop}%
\bibitem [{\citenamefont {Ohki}\ \emph {et~al.}(2004)\citenamefont {Ohki},
  \citenamefont {Nakagawa}, \citenamefont {Hirano}, \citenamefont {Hashimoto},
  \citenamefont {Menezes}, \citenamefont {Jokura}, \citenamefont {Uenohara},
  \citenamefont {Sato}, \citenamefont {Saito}, \citenamefont {Shirane} \emph
  {et~al.}}]{ohki2004experimental}%
  \BibitemOpen
  \bibfield  {author} {\bibinfo {author} {\bibfnamefont {T.}~\bibnamefont
  {Ohki}}, \bibinfo {author} {\bibfnamefont {A.}~\bibnamefont {Nakagawa}},
  \bibinfo {author} {\bibfnamefont {T.}~\bibnamefont {Hirano}}, \bibinfo
  {author} {\bibfnamefont {T.}~\bibnamefont {Hashimoto}}, \bibinfo {author}
  {\bibfnamefont {V.}~\bibnamefont {Menezes}}, \bibinfo {author} {\bibfnamefont
  {H.}~\bibnamefont {Jokura}}, \bibinfo {author} {\bibfnamefont
  {H.}~\bibnamefont {Uenohara}}, \bibinfo {author} {\bibfnamefont
  {Y.}~\bibnamefont {Sato}}, \bibinfo {author} {\bibfnamefont {T.}~\bibnamefont
  {Saito}}, \bibinfo {author} {\bibfnamefont {R.}~\bibnamefont {Shirane}},
  \emph {et~al.},\ }\href {\doibase 10.1002/lsm.20021} {\bibfield  {journal}
  {\bibinfo  {journal} {Lasers in surgery and medicine}\ }\textbf {\bibinfo
  {volume} {34}},\ \bibinfo {pages} {227} (\bibinfo {year} {2004})}\BibitemShut
  {NoStop}%
\bibitem [{\citenamefont {Duocastella}\ \emph {et~al.}(2008)\citenamefont
  {Duocastella}, \citenamefont {Fern{\'a}ndez-Pradas}, \citenamefont {Serra},\
  and\ \citenamefont {Morenza}}]{duocastella2008jet}%
  \BibitemOpen
  \bibfield  {author} {\bibinfo {author} {\bibfnamefont {M.}~\bibnamefont
  {Duocastella}}, \bibinfo {author} {\bibfnamefont {J.}~\bibnamefont
  {Fern{\'a}ndez-Pradas}}, \bibinfo {author} {\bibfnamefont {P.}~\bibnamefont
  {Serra}}, \ and\ \bibinfo {author} {\bibfnamefont {J.}~\bibnamefont
  {Morenza}},\ }\href {\doibase 10.1007/s00339-008-4781-y} {\bibfield
  {journal} {\bibinfo  {journal} {Applied Physics A}\ }\textbf {\bibinfo
  {volume} {93}},\ \bibinfo {pages} {453} (\bibinfo {year} {2008})}\BibitemShut
  {NoStop}%
\bibitem [{\citenamefont {Unger}\ \emph {et~al.}(2011)\citenamefont {Unger},
  \citenamefont {Gruene}, \citenamefont {Koch}, \citenamefont {Koch},\ and\
  \citenamefont {Chichkov}}]{unger2011time}%
  \BibitemOpen
  \bibfield  {author} {\bibinfo {author} {\bibfnamefont {C.}~\bibnamefont
  {Unger}}, \bibinfo {author} {\bibfnamefont {M.}~\bibnamefont {Gruene}},
  \bibinfo {author} {\bibfnamefont {L.}~\bibnamefont {Koch}}, \bibinfo {author}
  {\bibfnamefont {J.}~\bibnamefont {Koch}}, \ and\ \bibinfo {author}
  {\bibfnamefont {B.~N.}\ \bibnamefont {Chichkov}},\ }\href {\doibase
  10.1007/s00339-010-6030-4} {\bibfield  {journal} {\bibinfo  {journal}
  {Applied Physics A}\ }\textbf {\bibinfo {volume} {103}},\ \bibinfo {pages}
  {271} (\bibinfo {year} {2011})}\BibitemShut {NoStop}%
\bibitem [{\citenamefont {Brown}\ \emph {et~al.}(2011)\citenamefont {Brown},
  \citenamefont {Kattamis},\ and\ \citenamefont {Arnold}}]{brown2011time}%
  \BibitemOpen
  \bibfield  {author} {\bibinfo {author} {\bibfnamefont {M.~S.}\ \bibnamefont
  {Brown}}, \bibinfo {author} {\bibfnamefont {N.~T.}\ \bibnamefont {Kattamis}},
  \ and\ \bibinfo {author} {\bibfnamefont {C.~B.}\ \bibnamefont {Arnold}},\
  }\href {\doibase 10.1007/s10404-011-0787-4} {\bibfield  {journal} {\bibinfo
  {journal} {Microfluidics and nanofluidics}\ }\textbf {\bibinfo {volume}
  {11}},\ \bibinfo {pages} {199} (\bibinfo {year} {2011})}\BibitemShut
  {NoStop}%
\bibitem [{\citenamefont {Brasz}\ \emph {et~al.}(2015)\citenamefont {Brasz},
  \citenamefont {Yang},\ and\ \citenamefont {Arnold}}]{brasz2015tilting}%
  \BibitemOpen
  \bibfield  {author} {\bibinfo {author} {\bibfnamefont {C.~F.}\ \bibnamefont
  {Brasz}}, \bibinfo {author} {\bibfnamefont {J.~H.}\ \bibnamefont {Yang}}, \
  and\ \bibinfo {author} {\bibfnamefont {C.~B.}\ \bibnamefont {Arnold}},\
  }\href {\doibase 10.1007/s10404-014-1429-4} {\bibfield  {journal} {\bibinfo
  {journal} {Microfluidics and Nanofluidics}\ }\textbf {\bibinfo {volume}
  {18}},\ \bibinfo {pages} {185} (\bibinfo {year} {2015})}\BibitemShut
  {NoStop}%
\bibitem [{\citenamefont {Klein}\ \emph {et~al.}(2020)\citenamefont {Klein},
  \citenamefont {Kurilovich}, \citenamefont {Lhuissier}, \citenamefont
  {Versolato}, \citenamefont {Lohse}, \citenamefont {Villermaux},\ and\
  \citenamefont {Gelderblom}}]{klein2020drop}%
  \BibitemOpen
  \bibfield  {author} {\bibinfo {author} {\bibfnamefont {A.~L.}\ \bibnamefont
  {Klein}}, \bibinfo {author} {\bibfnamefont {D.}~\bibnamefont {Kurilovich}},
  \bibinfo {author} {\bibfnamefont {H.}~\bibnamefont {Lhuissier}}, \bibinfo
  {author} {\bibfnamefont {O.~O.}\ \bibnamefont {Versolato}}, \bibinfo {author}
  {\bibfnamefont {D.}~\bibnamefont {Lohse}}, \bibinfo {author} {\bibfnamefont
  {E.}~\bibnamefont {Villermaux}}, \ and\ \bibinfo {author} {\bibfnamefont
  {H.}~\bibnamefont {Gelderblom}},\ }\href {\doibase 10.1017/jfm.2020.197}
  {\bibfield  {journal} {\bibinfo  {journal} {Journal of Fluid Mechanics}\
  }\textbf {\bibinfo {volume} {893}},\ \bibinfo {pages} {A7} (\bibinfo {year}
  {2020})}\BibitemShut {NoStop}%
\bibitem [{\citenamefont {Thoroddsen}\ \emph {et~al.}(2009)\citenamefont
  {Thoroddsen}, \citenamefont {Takehara}, \citenamefont {Etoh},\ and\
  \citenamefont {Ohl}}]{thoroddsen2009spray}%
  \BibitemOpen
  \bibfield  {author} {\bibinfo {author} {\bibfnamefont {S.~T.}\ \bibnamefont
  {Thoroddsen}}, \bibinfo {author} {\bibfnamefont {K.}~\bibnamefont
  {Takehara}}, \bibinfo {author} {\bibfnamefont {T.}~\bibnamefont {Etoh}}, \
  and\ \bibinfo {author} {\bibfnamefont {C.-D.}\ \bibnamefont {Ohl}},\ }\href
  {\doibase 10.1063/1.3253394} {\bibfield  {journal} {\bibinfo  {journal}
  {Physics of Fluids}\ }\textbf {\bibinfo {volume} {21}},\ \bibinfo {pages}
  {112101} (\bibinfo {year} {2009})}\BibitemShut {NoStop}%
\bibitem [{\citenamefont {Li}\ \emph {et~al.}(2019)\citenamefont {Li},
  \citenamefont {Wang},\ and\ \citenamefont {Wang}}]{li2019cavitation}%
  \BibitemOpen
  \bibfield  {author} {\bibinfo {author} {\bibfnamefont {H.}~\bibnamefont
  {Li}}, \bibinfo {author} {\bibfnamefont {J.}~\bibnamefont {Wang}}, \ and\
  \bibinfo {author} {\bibfnamefont {Y.}~\bibnamefont {Wang}},\ }in\ \href
  {\doibase 10.1115/AJKFluids2019-5360} {\emph {\bibinfo {booktitle}
  {Computational Fluid Dynamics}}},\ \bibinfo {series} {Fluids Engineering
  Division Summer Meeting}, Vol.~\bibinfo {volume} {2}\ (\bibinfo {year}
  {2019})\ \bibinfo {note} {{V002T02A042}}\BibitemShut {NoStop}%
\bibitem [{\citenamefont {Zeng}\ \emph {et~al.}(2018)\citenamefont {Zeng},
  \citenamefont {Gonzalez-Avila}, \citenamefont {Ten~Voorde},\ and\
  \citenamefont {Ohl}}]{zeng2018jetting}%
  \BibitemOpen
  \bibfield  {author} {\bibinfo {author} {\bibfnamefont {Q.}~\bibnamefont
  {Zeng}}, \bibinfo {author} {\bibfnamefont {S.~R.}\ \bibnamefont
  {Gonzalez-Avila}}, \bibinfo {author} {\bibfnamefont {S.}~\bibnamefont
  {Ten~Voorde}}, \ and\ \bibinfo {author} {\bibfnamefont {C.-D.}\ \bibnamefont
  {Ohl}},\ }\href {\doibase 10.1017/jfm.2018.284} {\bibfield  {journal}
  {\bibinfo  {journal} {Journal of Fluid Mechanics}\ }\textbf {\bibinfo
  {volume} {846}},\ \bibinfo {pages} {916} (\bibinfo {year}
  {2018})}\BibitemShut {NoStop}%
\bibitem [{\citenamefont {Avila}\ and\ \citenamefont
  {Ohl}(2016)}]{avila2016fragmentation}%
  \BibitemOpen
  \bibfield  {author} {\bibinfo {author} {\bibfnamefont {S.~R.~G.}\
  \bibnamefont {Avila}}\ and\ \bibinfo {author} {\bibfnamefont {C.-D.}\
  \bibnamefont {Ohl}},\ }\href {\doibase 10.1017/jfm.2016.583} {\bibfield
  {journal} {\bibinfo  {journal} {Journal of Fluid Mechanics}\ }\textbf
  {\bibinfo {volume} {805}},\ \bibinfo {pages} {551} (\bibinfo {year}
  {2016})}\BibitemShut {NoStop}%
\bibitem [{\citenamefont {Plesset}(1954)}]{plesset1954stability}%
  \BibitemOpen
  \bibfield  {author} {\bibinfo {author} {\bibfnamefont {M.}~\bibnamefont
  {Plesset}},\ }\href {\doibase 10.1063/1.1721529} {\bibfield  {journal}
  {\bibinfo  {journal} {Journal of Applied Physics}\ }\textbf {\bibinfo
  {volume} {25}},\ \bibinfo {pages} {96} (\bibinfo {year} {1954})}\BibitemShut
  {NoStop}%
\bibitem [{\citenamefont {Prosperetti}(1977)}]{prosperetti1977viscous}%
  \BibitemOpen
  \bibfield  {author} {\bibinfo {author} {\bibfnamefont {A.}~\bibnamefont
  {Prosperetti}},\ }\href {\doibase 10.1090/qam/99652} {\bibfield  {journal}
  {\bibinfo  {journal} {Quarterly of Applied Mathematics}\ }\textbf {\bibinfo
  {volume} {34}},\ \bibinfo {pages} {339} (\bibinfo {year} {1977})}\BibitemShut
  {NoStop}%
\bibitem [{\citenamefont {Versolato}\ \emph {et~al.}(2022)\citenamefont
  {Versolato}, \citenamefont {Sheil}, \citenamefont {Witte}, \citenamefont
  {Ubachs},\ and\ \citenamefont {Hoekstra}}]{versolato2022microdroplet}%
  \BibitemOpen
  \bibfield  {author} {\bibinfo {author} {\bibfnamefont {O.}~\bibnamefont
  {Versolato}}, \bibinfo {author} {\bibfnamefont {J.}~\bibnamefont {Sheil}},
  \bibinfo {author} {\bibfnamefont {S.}~\bibnamefont {Witte}}, \bibinfo
  {author} {\bibfnamefont {W.}~\bibnamefont {Ubachs}}, \ and\ \bibinfo {author}
  {\bibfnamefont {R.}~\bibnamefont {Hoekstra}},\ }\href {\doibase
  10.1088/2040-8986/ac5a7e} {\bibfield  {journal} {\bibinfo  {journal} {Journal
  of Optics}\ } (\bibinfo {year} {2022}),\ 10.1088/2040-8986/ac5a7e},\ \bibinfo
  {note} {accepted for publication}\BibitemShut {NoStop}%
\bibitem [{\citenamefont {Krivokorytov}\ \emph {et~al.}(2018)\citenamefont
  {Krivokorytov}, \citenamefont {Zeng}, \citenamefont {Lakatosh}, \citenamefont
  {Vinokhodov}, \citenamefont {Sidelnikov}, \citenamefont {Kompanets},
  \citenamefont {Krivtsun}, \citenamefont {Koshelev}, \citenamefont {Ohl},\
  and\ \citenamefont {Medvedev}}]{krivokorytov:2018}%
  \BibitemOpen
  \bibfield  {author} {\bibinfo {author} {\bibfnamefont {M.}~\bibnamefont
  {Krivokorytov}}, \bibinfo {author} {\bibfnamefont {Q.}~\bibnamefont {Zeng}},
  \bibinfo {author} {\bibfnamefont {B.}~\bibnamefont {Lakatosh}}, \bibinfo
  {author} {\bibfnamefont {A.~Y.}\ \bibnamefont {Vinokhodov}}, \bibinfo
  {author} {\bibfnamefont {Y.~V.}\ \bibnamefont {Sidelnikov}}, \bibinfo
  {author} {\bibfnamefont {V.}~\bibnamefont {Kompanets}}, \bibinfo {author}
  {\bibfnamefont {V.}~\bibnamefont {Krivtsun}}, \bibinfo {author}
  {\bibfnamefont {K.}~\bibnamefont {Koshelev}}, \bibinfo {author}
  {\bibfnamefont {C.}~\bibnamefont {Ohl}}, \ and\ \bibinfo {author}
  {\bibfnamefont {V.}~\bibnamefont {Medvedev}},\ }\href {\doibase
  10.1038/s41598-017-19140-w} {\bibfield  {journal} {\bibinfo  {journal}
  {Scientific reports}\ }\textbf {\bibinfo {volume} {8}},\ \bibinfo {pages}
  {597} (\bibinfo {year} {2018})}\BibitemShut {NoStop}%
\bibitem [{\citenamefont {Grigoryev}\ \emph {et~al.}(2018)\citenamefont
  {Grigoryev}, \citenamefont {Lakatosh}, \citenamefont {Krivokorytov},
  \citenamefont {Zhakhovsky}, \citenamefont {Dyachkov}, \citenamefont
  {Ilnitsky}, \citenamefont {Migdal}, \citenamefont {Inogamov}, \citenamefont
  {Vinokhodov}, \citenamefont {Kompanets} \emph
  {et~al.}}]{grigoryev2018expansion}%
  \BibitemOpen
  \bibfield  {author} {\bibinfo {author} {\bibfnamefont {S.~Y.}\ \bibnamefont
  {Grigoryev}}, \bibinfo {author} {\bibfnamefont {B.}~\bibnamefont {Lakatosh}},
  \bibinfo {author} {\bibfnamefont {M.}~\bibnamefont {Krivokorytov}}, \bibinfo
  {author} {\bibfnamefont {V.}~\bibnamefont {Zhakhovsky}}, \bibinfo {author}
  {\bibfnamefont {S.}~\bibnamefont {Dyachkov}}, \bibinfo {author}
  {\bibfnamefont {D.}~\bibnamefont {Ilnitsky}}, \bibinfo {author}
  {\bibfnamefont {K.}~\bibnamefont {Migdal}}, \bibinfo {author} {\bibfnamefont
  {N.}~\bibnamefont {Inogamov}}, \bibinfo {author} {\bibfnamefont {A.~Y.}\
  \bibnamefont {Vinokhodov}}, \bibinfo {author} {\bibfnamefont
  {V.}~\bibnamefont {Kompanets}},  \emph {et~al.},\ }\href {\doibase
  10.1103/PhysRevApplied.10.064009} {\bibfield  {journal} {\bibinfo  {journal}
  {Physical Review Applied}\ }\textbf {\bibinfo {volume} {10}},\ \bibinfo
  {pages} {064009} (\bibinfo {year} {2018})}\BibitemShut {NoStop}%
\bibitem [{\citenamefont {de~Faria~Pinto}\ \emph {et~al.}(2021)\citenamefont
  {de~Faria~Pinto}, \citenamefont {Mathijssen}, \citenamefont {Meijer},
  \citenamefont {Zhang}, \citenamefont {Bayerle}, \citenamefont {Kurilovich},
  \citenamefont {Versolato}, \citenamefont {Eikema},\ and\ \citenamefont
  {Witte}}]{de2021cylindrically}%
  \BibitemOpen
  \bibfield  {author} {\bibinfo {author} {\bibfnamefont {T.}~\bibnamefont
  {de~Faria~Pinto}}, \bibinfo {author} {\bibfnamefont {J.}~\bibnamefont
  {Mathijssen}}, \bibinfo {author} {\bibfnamefont {R.}~\bibnamefont {Meijer}},
  \bibinfo {author} {\bibfnamefont {H.}~\bibnamefont {Zhang}}, \bibinfo
  {author} {\bibfnamefont {A.}~\bibnamefont {Bayerle}}, \bibinfo {author}
  {\bibfnamefont {D.}~\bibnamefont {Kurilovich}}, \bibinfo {author}
  {\bibfnamefont {O.~O.}\ \bibnamefont {Versolato}}, \bibinfo {author}
  {\bibfnamefont {K.~S.}\ \bibnamefont {Eikema}}, \ and\ \bibinfo {author}
  {\bibfnamefont {S.}~\bibnamefont {Witte}},\ }\href {\doibase
  10.1007/s00339-020-04207-9} {\bibfield  {journal} {\bibinfo  {journal}
  {Applied Physics A}\ }\textbf {\bibinfo {volume} {127}},\ \bibinfo {pages}
  {1} (\bibinfo {year} {2021})}\BibitemShut {NoStop}%
\bibitem [{\citenamefont {Vinokhodov}\ \emph {et~al.}(2016)\citenamefont
  {Vinokhodov}, \citenamefont {Koshelev}, \citenamefont {Krivtsun},
  \citenamefont {Krivokorytov}, \citenamefont {Sidelnikov}, \citenamefont
  {Medvedev}, \citenamefont {Kompanets}, \citenamefont {Melnikov},\ and\
  \citenamefont {Chekalin}}]{vinokhodov_dropgen}%
  \BibitemOpen
  \bibfield  {author} {\bibinfo {author} {\bibfnamefont {A.~Y.}\ \bibnamefont
  {Vinokhodov}}, \bibinfo {author} {\bibfnamefont {K.~N.}\ \bibnamefont
  {Koshelev}}, \bibinfo {author} {\bibfnamefont {V.~M.}\ \bibnamefont
  {Krivtsun}}, \bibinfo {author} {\bibfnamefont {M.~S.}\ \bibnamefont
  {Krivokorytov}}, \bibinfo {author} {\bibfnamefont {Y.~V.}\ \bibnamefont
  {Sidelnikov}}, \bibinfo {author} {\bibfnamefont {V.~V.}\ \bibnamefont
  {Medvedev}}, \bibinfo {author} {\bibfnamefont {V.~O.}\ \bibnamefont
  {Kompanets}}, \bibinfo {author} {\bibfnamefont {A.~A.}\ \bibnamefont
  {Melnikov}}, \ and\ \bibinfo {author} {\bibfnamefont {S.~V.}\ \bibnamefont
  {Chekalin}},\ }\href {\doibase 10.1070/QE2016v046n01ABEH015867} {\bibfield
  {journal} {\bibinfo  {journal} {Quantum Electronics}\ }\textbf {\bibinfo
  {volume} {46}},\ \bibinfo {pages} {23} (\bibinfo {year} {2016})}\BibitemShut
  {NoStop}%
\bibitem [{\citenamefont {Zhakhovskii}\ \emph {et~al.}(2005)\citenamefont
  {Zhakhovskii}, \citenamefont {Nishihara}, \citenamefont {Fukuda},
  \citenamefont {Shimojo}, \citenamefont {Akiyama}, \citenamefont {Miyanaga}
  \emph {et~al.}}]{Zhakhovskii:2005}%
  \BibitemOpen
  \bibfield  {author} {\bibinfo {author} {\bibfnamefont {V.}~\bibnamefont
  {Zhakhovskii}}, \bibinfo {author} {\bibfnamefont {K.}~\bibnamefont
  {Nishihara}}, \bibinfo {author} {\bibfnamefont {Y.}~\bibnamefont {Fukuda}},
  \bibinfo {author} {\bibfnamefont {S.}~\bibnamefont {Shimojo}}, \bibinfo
  {author} {\bibfnamefont {T.}~\bibnamefont {Akiyama}}, \bibinfo {author}
  {\bibfnamefont {S.}~\bibnamefont {Miyanaga}},  \emph {et~al.},\ }in\ \href
  {\doibase 10.1109/CCGRID.2005.1558650} {\emph {\bibinfo {booktitle} {CCGrid
  2005. IEEE International Symposium on Cluster Computing and the Grid,
  2005.}}},\ Vol.~\bibinfo {volume} {2}\ (\bibinfo {organization} {IEEE},\
  \bibinfo {year} {2005})\ pp.\ \bibinfo {pages} {848--854}\BibitemShut
  {NoStop}%
\bibitem [{\citenamefont {Egorova}\ \emph {et~al.}(2019)\citenamefont
  {Egorova}, \citenamefont {Dyachkov}, \citenamefont {Parshikov},\ and\
  \citenamefont {Zhakhovsky}}]{Egorova:2019}%
  \BibitemOpen
  \bibfield  {author} {\bibinfo {author} {\bibfnamefont {M.~S.}\ \bibnamefont
  {Egorova}}, \bibinfo {author} {\bibfnamefont {S.~A.}\ \bibnamefont
  {Dyachkov}}, \bibinfo {author} {\bibfnamefont {A.~N.}\ \bibnamefont
  {Parshikov}}, \ and\ \bibinfo {author} {\bibfnamefont {V.~V.}\ \bibnamefont
  {Zhakhovsky}},\ }\href {\doibase 10.1016/j.cpc.2018.07.019} {\bibfield
  {journal} {\bibinfo  {journal} {Computer Physics Communications}\ }\textbf
  {\bibinfo {volume} {234}},\ \bibinfo {pages} {112} (\bibinfo {year}
  {2019})}\BibitemShut {NoStop}%
\bibitem [{\citenamefont {Zhakhovskii}\ \emph {et~al.}(2009)\citenamefont
  {Zhakhovskii}, \citenamefont {Inogamov}, \citenamefont {Petrov},
  \citenamefont {Ashitkov},\ and\ \citenamefont
  {Nishihara}}]{Zhakhovskii:2009}%
  \BibitemOpen
  \bibfield  {author} {\bibinfo {author} {\bibfnamefont {V.~V.}\ \bibnamefont
  {Zhakhovskii}}, \bibinfo {author} {\bibfnamefont {N.~A.}\ \bibnamefont
  {Inogamov}}, \bibinfo {author} {\bibfnamefont {Y.~V.}\ \bibnamefont
  {Petrov}}, \bibinfo {author} {\bibfnamefont {S.~I.}\ \bibnamefont
  {Ashitkov}}, \ and\ \bibinfo {author} {\bibfnamefont {K.}~\bibnamefont
  {Nishihara}},\ }\href {\doibase 10.1016/j.apsusc.2009.04.082} {\bibfield
  {journal} {\bibinfo  {journal} {Appl. Surf. Sci.}\ }\textbf {\bibinfo
  {volume} {255}},\ \bibinfo {pages} {9592} (\bibinfo {year}
  {2009})}\BibitemShut {NoStop}%
\bibitem [{\citenamefont {Zhukhovitskii}\ and\ \citenamefont
  {Zhakhovsky}(2020)}]{Zhukhovitskii:2020}%
  \BibitemOpen
  \bibfield  {author} {\bibinfo {author} {\bibfnamefont {D.~I.}\ \bibnamefont
  {Zhukhovitskii}}\ and\ \bibinfo {author} {\bibfnamefont {V.~V.}\ \bibnamefont
  {Zhakhovsky}},\ }\href {\doibase 10.1063/5.0010156} {\bibfield  {journal}
  {\bibinfo  {journal} {The Journal of Chemical Physics}\ }\textbf {\bibinfo
  {volume} {152}},\ \bibinfo {pages} {224705} (\bibinfo {year}
  {2020})}\BibitemShut {NoStop}%
\bibitem [{\citenamefont {Zhakhovsky}()}]{Zhakhovsky:EAM-project}%
  \BibitemOpen
  \bibfield  {author} {\bibinfo {author} {\bibfnamefont {V.~V.}\ \bibnamefont
  {Zhakhovsky}},\ }\href@noop {} {}\bibinfo {note} {Tabulated {EAM} potential
  for liquid tin can be downloaded from
  \href{https://www.researchgate.net/project/Development-of-interatomic-EAM-potentials}
  {www.researchgate.net/project/Development-of-interatomic-EAM-potentials}}\BibitemShut
  {NoStop}%
\bibitem [{\citenamefont {Volkov}\ and\ \citenamefont
  {Sibilev}(1981)}]{volkov1981investigation}%
  \BibitemOpen
  \bibfield  {author} {\bibinfo {author} {\bibfnamefont {K.}~\bibnamefont
  {Volkov}}\ and\ \bibinfo {author} {\bibfnamefont {V.}~\bibnamefont
  {Sibilev}},\ }\href {\doibase 10.1007/BF00906269} {\bibfield  {journal}
  {\bibinfo  {journal} {Journal of Applied Mechanics and Technical Physics}\
  }\textbf {\bibinfo {volume} {22}},\ \bibinfo {pages} {551} (\bibinfo {year}
  {1981})}\BibitemShut {NoStop}%
\bibitem [{\citenamefont {Trunin}\ \emph {et~al.}(1995)\citenamefont {Trunin},
  \citenamefont {Zhernokletov}, \citenamefont {Kuznetsov},\ and\ \citenamefont
  {Shutov}}]{trunin1995dynamic}%
  \BibitemOpen
  \bibfield  {author} {\bibinfo {author} {\bibfnamefont {R.~F.}\ \bibnamefont
  {Trunin}}, \bibinfo {author} {\bibfnamefont {M.}~\bibnamefont
  {Zhernokletov}}, \bibinfo {author} {\bibfnamefont {N.}~\bibnamefont
  {Kuznetsov}}, \ and\ \bibinfo {author} {\bibfnamefont {V.}~\bibnamefont
  {Shutov}},\ }\href@noop {} {\bibfield  {journal} {\bibinfo  {journal}
  {Teplofizika Vysokikh Temperatur}\ }\textbf {\bibinfo {volume} {33}},\
  \bibinfo {pages} {222} (\bibinfo {year} {1995})}\BibitemShut {NoStop}%
\bibitem [{\citenamefont {Rowlinson}\ and\ \citenamefont
  {Widom}(1982)}]{rowlinson1982molecular}%
  \BibitemOpen
  \bibfield  {author} {\bibinfo {author} {\bibfnamefont {J.}~\bibnamefont
  {Rowlinson}}\ and\ \bibinfo {author} {\bibfnamefont {B.}~\bibnamefont
  {Widom}},\ }\href@noop {} {\emph {\bibinfo {title} {Molecular Theory of
  Capillarity}}}\ (\bibinfo  {publisher} {Clarendon Press, Oxford},\ \bibinfo
  {year} {1982})\ p.\ \bibinfo {pages} {586}\BibitemShut {NoStop}%
\bibitem [{\citenamefont {Kononenko}\ \emph {et~al.}(1972)\citenamefont
  {Kononenko}, \citenamefont {Sukhman},\ and\ \citenamefont
  {Yatsenko}}]{kononenko1972surface}%
  \BibitemOpen
  \bibfield  {author} {\bibinfo {author} {\bibfnamefont {V.~I.}\ \bibnamefont
  {Kononenko}}, \bibinfo {author} {\bibfnamefont {A.~L.}\ \bibnamefont
  {Sukhman}}, \ and\ \bibinfo {author} {\bibfnamefont {S.~P.}\ \bibnamefont
  {Yatsenko}},\ }\href@noop {} {\bibfield  {journal} {\bibinfo  {journal}
  {Russian Journal of Physical Chemistry}\ }\textbf {\bibinfo {volume} {46}},\
  \bibinfo {pages} {911} (\bibinfo {year} {1972})}\BibitemShut {NoStop}%
\bibitem [{\citenamefont {Passerone}\ \emph {et~al.}(1990)\citenamefont
  {Passerone}, \citenamefont {Ricci},\ and\ \citenamefont
  {Sangiorgi}}]{passerone1990influence}%
  \BibitemOpen
  \bibfield  {author} {\bibinfo {author} {\bibfnamefont {A.}~\bibnamefont
  {Passerone}}, \bibinfo {author} {\bibfnamefont {E.}~\bibnamefont {Ricci}}, \
  and\ \bibinfo {author} {\bibfnamefont {R.}~\bibnamefont {Sangiorgi}},\ }\href
  {\doibase 10.1007/BF00581083} {\bibfield  {journal} {\bibinfo  {journal}
  {Journal of materials science}\ }\textbf {\bibinfo {volume} {25}},\ \bibinfo
  {pages} {4266} (\bibinfo {year} {1990})}\BibitemShut {NoStop}%
\bibitem [{\citenamefont {Fima}\ \emph {et~al.}(2010)\citenamefont {Fima},
  \citenamefont {Nowak},\ and\ \citenamefont {Sobczak}}]{fima2010effect}%
  \BibitemOpen
  \bibfield  {author} {\bibinfo {author} {\bibfnamefont {P.}~\bibnamefont
  {Fima}}, \bibinfo {author} {\bibfnamefont {R.}~\bibnamefont {Nowak}}, \ and\
  \bibinfo {author} {\bibfnamefont {N.}~\bibnamefont {Sobczak}},\ }\href
  {\doibase 10.1007/s10853-009-3973-y} {\bibfield  {journal} {\bibinfo
  {journal} {Journal of materials science}\ }\textbf {\bibinfo {volume} {45}},\
  \bibinfo {pages} {2009} (\bibinfo {year} {2010})}\BibitemShut {NoStop}%
\bibitem [{\citenamefont {Palik}(1998)}]{palik1998handbook}%
  \BibitemOpen
  \bibfield  {author} {\bibinfo {author} {\bibfnamefont {E.~D.}\ \bibnamefont
  {Palik}},\ }\href@noop {} {\emph {\bibinfo {title} {Handbook of optical
  constants of solids}}},\ \bibinfo {series} {Academic Press handbook series},
  Vol.~\bibinfo {volume} {3}\ (\bibinfo  {publisher} {Academic press},\
  \bibinfo {year} {1998})\BibitemShut {NoStop}%
\bibitem [{\citenamefont {Petrakian}\ \emph {et~al.}(1980)\citenamefont
  {Petrakian}, \citenamefont {Cathers}, \citenamefont {Parks}, \citenamefont
  {MacRae}, \citenamefont {Callcott},\ and\ \citenamefont
  {Arakawa}}]{petrakian1980optical}%
  \BibitemOpen
  \bibfield  {author} {\bibinfo {author} {\bibfnamefont {J.}~\bibnamefont
  {Petrakian}}, \bibinfo {author} {\bibfnamefont {A.}~\bibnamefont {Cathers}},
  \bibinfo {author} {\bibfnamefont {J.}~\bibnamefont {Parks}}, \bibinfo
  {author} {\bibfnamefont {R.}~\bibnamefont {MacRae}}, \bibinfo {author}
  {\bibfnamefont {T.}~\bibnamefont {Callcott}}, \ and\ \bibinfo {author}
  {\bibfnamefont {E.}~\bibnamefont {Arakawa}},\ }\href {\doibase
  10.1103/PhysRevB.21.3043} {\bibfield  {journal} {\bibinfo  {journal}
  {Physical Review B}\ }\textbf {\bibinfo {volume} {21}},\ \bibinfo {pages}
  {3043} (\bibinfo {year} {1980})}\BibitemShut {NoStop}%
\end{thebibliography}
\end{document}